\documentclass[aps,prd,superscriptaddress,10]{revtex4-2}
\usepackage{graphicx,float,wrapfig,subfigure}
\usepackage{amsfonts,amsmath,amssymb,amstext}
\usepackage{latexsym}
\usepackage{bm}
\usepackage{color}
\usepackage[normalem]{ulem}
\usepackage{MnSymbol}
\usepackage[colorlinks=true,linktocpage=true,linkcolor=blue,citecolor=blue]{hyperref}
\usepackage{slashed}
\usepackage{orcidlink}

\newcommand{\sign}{ \mbox{sign}}
\renewcommand{\Im}{\mbox{Im}}

\definecolor{red}{rgb}{0.7,0,0}
\definecolor{green}{rgb}{0,0.5,0}

\begin{document}

\title{Neutrino-antineutrino synchrotron emission from magnetized dense quark matter}

\author{Ritesh Ghosh\orcidlink{0000-0002-6740-7038}}
\email{Ritesh.Ghosh@asu.edu}
\affiliation{College of Integrative Sciences and Arts, Arizona State University, Mesa, Arizona 85212, USA}

\author{Igor A. Shovkovy\orcidlink{0000-0002-5230-6891}}
\email{igor.shovkovy@asu.edu}
\affiliation{College of Integrative Sciences and Arts, Arizona State University, Mesa, Arizona 85212, USA}
\affiliation{Department of Physics, Arizona State University, Tempe, Arizona 85287, USA}

\date{April 29, 2025}

\begin{abstract}
Using the Kadanoff-Baym formalism, we perform a detailed study of neutrino-antineutrino synchrotron emission from strongly magnetized, dense quark matter under conditions relevant to compact stars. Starting from an exact expression for the emission rate that fully accounts for Landau-level quantization of quarks, we derive an approximate formula applicable in the regime where quark chemical potentials are much larger than all other relevant energy scales. We demonstrate that the emission rate is largely controlled by a single dimensionless ratio between two low-energy scales: the Landau-level spacing at the Fermi surface, $|e_f B|/\mu_{f}$, and the temperature of the quark matter, $T$. When the ratio $|e_f B|/(\mu_{f} T)$ approaches zero, many closely spaced Landau levels contribute to the emission, but the total rate vanishes as $B\to 0$. In the opposite limit, where the ratio is large, the rate is dominated by transitions between adjacent levels and is exponentially suppressed due to Landau-level quantization, which limits the thermal activation of quarks near the Fermi surface. Our results show that, even in the presence of the strongest magnetic fields expected in compact stars, the synchrotron emission remains suppressed by more than 3 orders of magnitude compared to the direct Urca process. This implies that such emission is unlikely to play any substantial role in the cooling of magnetized quark stars, at least those made of unpaired quark matter phases.
\end{abstract} 

\maketitle

\section{Introduction}
\label{sec:Introduction}

The interiors of compact stars offer a unique natural laboratory for investigating the behavior of matter under extreme conditions \cite{Baym:2017whm}. At densities several times greater than the nuclear saturation density, the environment becomes favorable for the emergence of a deconfined phase of baryonic matter composed of up and down quarks \cite{Annala:2023cwx}. In magnetars, as a special class of compact stars, conditions become even more extreme due to the presence of ultrastrong magnetic fields \cite{Turolla:2015mwa,Kaspi:2017fwg}, which can reach  strengths of $10^{15}~\mbox{G}$ on the surface and potentially even higher in the stellar interior \cite{Lai_1991}.

Numerous theoretical efforts have focused on making testable predictions to support the existence of quark matter in the cores of compact stars. One promising direction involves investigating the underlying cooling mechanisms \cite{Yakovlev:2004iq,Page:2005fq,Page:2006ud}, which are usually dominated by neutrino emission \cite{Yakovlev:2000jp}. In an earlier study \cite{Ghosh:2025sjn}, we examined the impact of strong magnetic fields on neutrino emission rates. Using first-principles, field-theoretic methods, we computed both the energy and net longitudinal momentum carried away by neutrinos. The emission is dominated by the direct Urca processes, namely, electron capture by an up quark ($u + e^- \rightarrow d + \nu_e$) and the decay of a down quark ($d \rightarrow u + e^- + \bar{\nu}_e$). These same weak interaction processes serve as the primary cooling channel even in the absence of magnetic fields \cite{Iwamoto:1980eb,Iwamoto:1982zz}. Thus, it is not unexpected that magnetic fields have only a moderate impact on them under the physical conditions relevant to compact stars.

It is generally instructive to investigate whether certain processes are enabled exclusively by the presence of strong magnetic fields. One such process in magnetized quark matter is neutrino-antineutrino synchrotron emission, $q_f \rightarrow q_f + \nu_i + \bar{\nu}_i$, where $q_f$ denotes an up or down quark ($f=u,d$) and $\nu_i$ represents a neutrino of a given flavor ($i=e,\mu,\tau$). In the absence of a background magnetic field, such synchrotron emission is forbidden by energy-momentum conservation.

Previously, analogous processes have been extensively studied in various forms of magnetized matter in white dwarfs and neutron stars \cite{Yakovlev:1981AN,Kaminker:1992su,Kaminker:1993ey,Vidaurre:1995iv,Bezchastnov:1997ew}, where $\nu\bar{\nu}$ synchrotron emission primarily originates from electrons via $e^{-} \rightarrow e^{-} + \nu_i + \bar{\nu}_i$. In these stars, such emission has been found to be significant, and even dominant, within certain density and temperature regimes. (For the earliest attempts to study $\nu\bar{\nu}$ synchrotron emission, see also Refs.~\cite{Landstreet:1967zz,Canuto:1970}.) However, similar emission has not been thoroughly explored in the context of quark matter. From general scaling arguments \cite{Kaminker:1992su}, one might speculate that synchrotron and direct Urca processes could become comparable under certain conditions. Yet, such qualitative arguments alone are insufficient to establish the actual relevance of synchrotron emission in dense quark matter. A detailed and systematic analysis is, therefore, necessary to assess its physical impact under realistic astrophysical conditions.

In this work, we undertake a detailed study of $\nu \bar{\nu}$ synchrotron emission from magnetized dense quark matter. Besides calculating the emission rate explicitly, we provide an alternative derivation employing a novel and efficient field-theoretical technique. Specifically, we utilize the Kadanoff-Baym formalism to determine the evolution of the neutrino (antineutrino) distribution function driven by weak interactions, from which we subsequently extract the corresponding energy emission rate. Given that a small density of electrons is inherently present in quark matter, we include their contribution in our calculations as well. Contrary to initial expectations, our analysis reveals that synchrotron emission is significantly suppressed relative to direct Urca emission, even in the presence of the strongest permissible magnetic fields. As we argue, the suppression originates not only from the scaling behavior with respect to temperature and magnetic field strength, but also from a numerically small overall prefactor in the emission rate.

This paper is organized as follows. In Sec.~\ref{sec:formalism}, we present the theoretical framework based on the Kadanoff-Baym formalism for computing $ \nu \bar{\nu} $ synchrotron emission in magnetized quark matter. Within this framework, we first derive the exact result and then obtain a suitable approximate expression applicable in the high-density regime, where the quark chemical potentials are much larger than all other relevant energy scales. In this limit, we identify the temperature and the Landau-level spacing near the quark Fermi surface as the sole low-energy parameters governing the emission rate. In Sec.~\ref{sec:results}, we present numerical results for the emission rates and examine their dependence on key model parameters. For completeness, we also account for the partial contribution to synchrotron emission from electrons, which are inevitably present in electrically neutral, $\beta$-equilibrated quark matter. Finally, in Sec.~\ref{sec:Summary}, we summarize our findings, discuss their implications for compact stars, and outline directions for future research. The Appendix contains the derivation of the imaginary part of the vector $Z$-boson self-energy and its Lorentz contraction with the lepton tensor.

\section{General Formalism}
\label{sec:formalism}

Traditionally, calculations of neutrino-antineutrino synchrotron emission begin with the transition amplitude for the underlying tree-level electroweak process, shown in Fig.~\ref{fig:EW-process}(a), using the exact electron Landau-level wave functions in a magnetic field \cite{Yakovlev:1981AN,Kaminker:1992su}. In this study, however, we adopt an alternative approach, similar to those employed in Refs.~\cite{Sedrakian:1999jh,Schmitt:2005wg,Berdermann:2016mwt,Ghosh:2025sjn}, which utilizes the powerful Kadanoff-Baym formalism. Unlike the traditional method, this formalism relies on Green functions rather than the more cumbersome particle wave functions, providing a more streamlined and efficient theoretical framework.

Here we consider dense quark matter composed of the two lightest up and down quark flavors. In $\beta$-equilibrium, a small density of electrons must also be present to maintain charge neutrality. The condition of $\beta$-equilibrium imposes a relation among the chemical potentials of the particles, namely, $\mu_d = \mu_u + \mu_e$, where $\mu_d$, $\mu_u$, and $\mu_e$ denote the chemical potentials of down quarks, up quarks, and electrons, respectively.

In the case of synchrotron emission, quarks and electrons contribute independently. For clarity and simplicity, in this section we focus on the emission from a single quark flavor producing a neutrino-antineutrino of a single flavor. This approach streamlines the derivation while retaining generality. When applying the results to quark stars, the general expressions can be easily adapted to account for all quark flavors, all neutrino types, and additional contributions from electron emission. The latter part, as we will verify, is the same as in Ref.~\cite{Kaminker:1992su}.

\subsection{Neutrino-number production rate}
\label{subsec:d-f-neutrino}

The main component in our calculation is the pair of out-of-equilibrium neutrino Green functions, $G^{\lessgtr}_\nu(t,P)$, which evolve slowly over time due to ongoing electroweak interactions. These functions are directly related to the neutrino and antineutrino distribution functions and can be expressed in the following forms \cite{Sedrakian:1999jh,Schmitt:2005wg,Berdermann:2016mwt,Ghosh:2025sjn}:
\begin{eqnarray}
 i G^{<}_\nu(t,P) &=&-\frac{\pi}{p} \frac{1-\gamma_5}{2} (\gamma^\lambda P_{\lambda}+\mu_\nu \gamma_0) \left\{f_\nu(t,\bm{p})\delta(p_{0}+\mu_\nu-p)-\left[1-f_{\bar \nu}(t,-\bm{p})\right]\delta(p_{0}+\mu_\nu+p)\right\} ,  \label{G-less}\\
 i G^{>}_\nu(t,P)& =&\frac{\pi}{p} \frac{1-\gamma_5}{2} (\gamma^\lambda P_{\lambda}+\mu_\nu \gamma_0) \left\{[1-f_\nu(t,\bm{p})]\delta(p_{0}+\mu_\nu-p)-f_{\bar \nu}(t,-\bm{p})\delta(p_{0}+\mu_\nu+p)\right\} .
 \label{G-gtr}
\end{eqnarray}
Here, by definition, $P = (p_{0}, \bm{p})$ denotes the four-momentum of the neutrino, with $p = |\bm{p}|$ representing the magnitude of its three-momentum. To account for the purely left-handed nature of neutrinos, the chirality projectors $(1 - \gamma_5)/2$ are included  in Eqs.~(\ref{G-less}) and (\ref{G-gtr}). Since we assume no neutrino trapping in quark matter, the neutrino chemical potential is set to zero, i.e., $\mu_\nu = 0$ throughout this study. The absence of neutrino trapping also implies that the neutrino distribution functions themselves are approximately vanishing, although their time derivatives are not.

%%%%%%%%%% FIGURE 1 %%%%%%%%%% 
\begin{figure}[t]
\centering
  \subfigure[]{\includegraphics[height=0.125\textwidth]{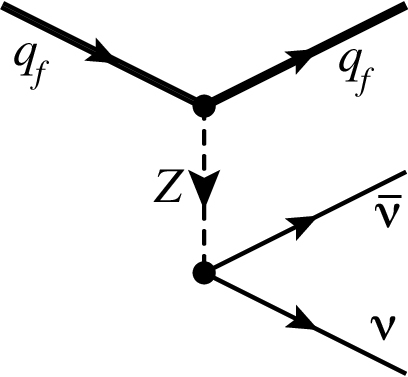}}
  \hspace{0.05\textwidth}
  \subfigure[]{\includegraphics[height=0.125\textwidth]{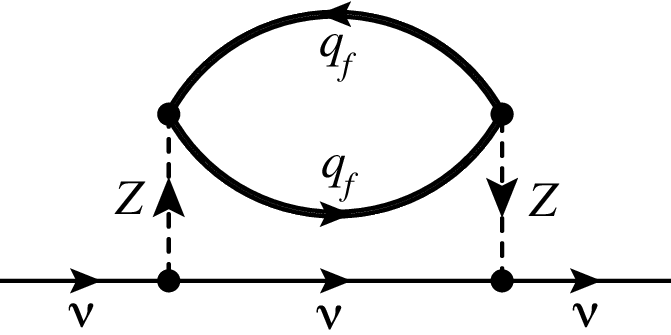}}
\caption{(a) Feynman diagram illustrating the neutral current interaction responsible for $\nu\bar{\nu}$ synchrotron emission in magnetized quark matter. (b) Neutrino self-energy diagram used in the calculation of the synchrotron emission rate.}
\label{fig:EW-process}
\end{figure}
%%%%%%%%%%%%%%%%%%%%

The Green functions satisfy the following equation:
\begin{equation}
 i \partial_t \mbox{Tr}[\gamma^0 G_\nu^{<} (t,P)] = -\mbox{Tr}[G_\nu^{>} (t,P)\Sigma^{<}_\nu(t,P)-\Sigma^{>}_\nu(t,P)G_\nu^{<} (t,P)],
 \label{KB-kinetic-eq}
\end{equation}
which serves as the field-theoretic analog of the kinetic equation for the (anti)neutrino distribution function. The effects of electroweak interactions are encoded in this equation through the neutrino self-energies, $\Sigma^{\lessgtr}_\nu(t,P)$.

To derive the relevant contribution to the neutrino self-energy, we use the Weinberg-Salam theory of weak interactions. Since the synchrotron emission is determined by the neutral-current interaction, the corresponding Lagrangian density reads \cite{Peskin:1995ev}
\begin{equation}
 \mathcal{L}=\frac{G_F }{\sqrt{2}} \bar \psi_\nu \gamma^\mu(1-\gamma_5) \psi_\nu \, \sum_{f}\bar{\psi_f} \gamma_\mu (c_V^f-c_A^f\gamma_5)\psi_f,
 \label{interaction-Lagrangian}
\end{equation}
where $G_F\approx 1.166\times 10^{-11}~\mbox{MeV}^{-2}$ is the Fermi coupling constant. The vector $(c_V^f)$ and axial-vector $(c_A^f)$ couplings appearing in the neutral currents of quarks and electrons are listed in Table~\ref{cVcA-coupling}.

\begin{table}[b]
\caption{\label{cVcA-coupling} Vector and axial-vector couplings in weak interactions. The weak mixing angle value is given by $\sin^2\theta_W \approx 0.231$.} 
\begin{ruledtabular}
\begin{tabular}{llllll}
Particle & $c_V^f$ & $c_A^f$ & $(c_V^f)^2+(c_A^f)^2$ & $(c_V^f)^2-(c_A^f)^2$ & $c_V^f c_A^f$ \\
\hline
$u\to u+\nu_i+\bar{\nu}_i$ & $\frac{1}{2}-\frac{4}{3}\sin^2\theta_W$ &  $\frac{1}{2}$ & 0.287 &  -0.213 & 0.096 \\
$d\to d+\nu_i+\bar{\nu}_i$ & $-\frac{1}{2}+\frac{2}{3}\sin^2\theta_W$ &  $-\frac{1}{2}$ &  0.370 & -0.130 & 0.173\\
$e^{-}\to e^{-}+\nu_{\mu,\tau}+\bar{\nu}_{\mu,\tau}$\footnote{Only neutral-current interaction contributes to synchrotron emission of muon and tau neutrinos.}
& $-\frac{1}{2}+2\sin^2\theta_W$ &  $-\frac{1}{2}$ & 0.251 & -0.249 & 0.019 \\
$e^{-}\to e^{-}+\nu_{e}+\bar{\nu}_{e}$\footnote{These are effective coupling constants accounting for the combined effect of both neutral- and charged-current interactions in $ \nu_{e} \bar{\nu}_{e} $ synchrotron emission \cite{Yakovlev:1981AN,Kaminker:1992su}.} 
& $\frac{1}{2}+2\sin^2\theta_W$ &  $\frac{1}{2}$ & 1.175 & 0.675 & 0.481 \\
\end{tabular} 
\end{ruledtabular} 
\end{table}

Starting from the interaction Lagrangian in Eq.~(\ref{interaction-Lagrangian}), we obtain the following expression for the neutrino self-energies:
\begin{equation}
 \Sigma^{\lessgtr}_\nu(P) =   i  \frac{G_F^2}{2} \int \frac{d^4 Q}{(2\pi)^4} \gamma^\delta (1-\gamma^5)
 G_\nu^{\lessgtr}(Q)\gamma^\sigma(1-\gamma^5){\Pi}^{\lessgtr}_{\delta\sigma}(P-Q) ,
 \label{Sigma-gtr}
\end{equation}
where ${\Pi}^{\lessgtr}_{\delta\sigma}(P-Q) $ are the lesser and greater $Z$-boson self-energies. The diagrammatic representation of the self-energy is shown in Fig.~\ref{fig:EW-process}(b).

By substituting the Green functions given in Eqs.~(\ref{G-less}) and (\ref{G-gtr}) into Eq.~(\ref{Sigma-gtr}) and taking into account that there is no neutrino trapping, i.e., that neutrino and antineutrino distribution functions are approximately zero, we derive the following result for the self-energies:
\begin{eqnarray}
\Sigma^{\lessgtr}_\nu(P) &=&   \frac{G_F^2}{4} \int \frac{d^4 Q}{(2\pi)^3 q_{0}} \gamma^\delta (1-\gamma^5) \gamma^\lambda Q_{\lambda} \frac{1-\gamma_5}{2}\delta(q_{0}\pm q)\gamma^\sigma(1-\gamma^5){\Pi}^{\lessgtr}_{\delta\sigma}(P-Q).
\end{eqnarray}
Finally, by inserting this expression into the kinetic equation (\ref{KB-kinetic-eq}), we derive the neutrino-number production rate 
\begin{eqnarray}
\frac{\partial f_\nu(t,\bm{p})}{\partial t}  &=& \frac{G_F^2}{4} \int \frac{d^3 \bm{q}}{(2\pi)^3 p_{0}q_{0}}  L^{\delta\sigma}(Q,P)
n_B(p_{0}+q_{0}) \Im \left[\Pi^R_{\delta\sigma}(P+Q)\right] ,
\label{d-f_nu-dt}
\end{eqnarray}
where $P_{\lambda} = (p_{0},\bm{p})$ and $Q_{\lambda} = (q_{0},\bm{q})$ are the four-momenta of the neutrino and antineutrino, respectively. (In the derivation, we changed the sign of the integration variable $\bm{q} \to - \bm{q}$.) Here, by definition, the timelike components of the momenta satisfy the on-shell conditions: $p_{0}=p$ and $q_{0}=q$. A similar expression can also be derived for the antineutrino-number production rate $\partial_t f_{\bar{\nu}}(t,\bm{p})$. One can show that the corresponding rate is identical to that for neutrinos, as given in Eq.~(\ref{d-f_nu-dt}). This is expected, of course, since the underlying weak processes produce neutrinos and antineutrinos in pairs, see Fig.~\ref{fig:EW-process}(a). 

In deriving the neutrino-number production rate in Eq.~(\ref{d-f_nu-dt}), we expressed the lesser and greater $Z$-boson self-energies in terms of the imaginary part of the retarded self-energy function \cite{2000tft..book.....L}, i.e.,
\begin{eqnarray}
 i\Pi_{\delta\sigma}^>(Q)&=& 2[1+n_B(q_0)] \Im \left[\Pi^R_{\delta\sigma}(Q)\right],\\
 i\Pi_{\delta\sigma}^<(Q)&=&2n_B(q_0)\Im \left[\Pi^R_{\delta\sigma}(Q)\right] ,
\end{eqnarray}
where $n_B(q_0) \equiv 1/\left[e^{q_0/T}-1\right]$ is the Bose-Einstein distribution function. 

We also introduced the shorthand notation $L^{\delta\sigma}(Q,P)$ for the lepton tensor, defined via the following Dirac trace:
\begin{eqnarray}
L^{\delta\sigma}(Q,P)&=&
\mbox{Tr}\left[(p_0\gamma^0-\bm{\gamma}\cdot \bm{p})  \gamma^\delta (1-\gamma^5) (q_0\gamma^0-\bm{\gamma}\cdot \bm{q})  \gamma^\sigma(1-\gamma^5)\right] \nonumber\\
&=& 8\left[ Q^\delta P^\sigma+P^\delta Q^\sigma-g^{\delta\sigma}(P^{\rho}Q_{\rho})
+i\epsilon^{\delta\sigma \kappa\rho}Q_\kappa P_\rho
\right] ,
\end{eqnarray}
where $\epsilon^{\delta\sigma \kappa\rho}$ denotes the conventional Levi-Civita tensor.

\subsection{Synchrotron emission rate}
\label{subsec:Synch-emission}

In terms of the neutrino and antineutrino distribution functions, the synchrotron emission rate is defined as
\begin{eqnarray}
\dot{\cal E}_\nu &=& \int \frac{d^3\bm{p} }{(2\pi)^3}  p_0 \frac{\partial f_\nu(t,\bm{p} )}{\partial t}
+ \int \frac{d^3\bm{q} }{(2\pi)^3}  q_0 \frac{\partial f_{\bar{\nu}}(t,\bm{q} )}{\partial t}.
\label{def-rate}
\end{eqnarray}
It is important to note that both neutrino and antineutrino energy emission rates are included in this definition. By making use of the neutrino-number production rate in Eq.~(\ref{d-f_nu-dt}) and taking into account that the result for antineutrinos is the same, we then derive
\begin{eqnarray}
\dot{\cal E}_\nu &=&  \frac{G_F^2}{4} \int \frac{d^3\bm{p} d^3 \bm{q}}{(2\pi)^6 q_0 p_0} 
 (p_0 +q_0) n_B( p_0+q_0) L^{\delta\sigma}(Q,P) \Im \left[\Pi^R_{\delta\sigma}(P+Q)\right].
\label{E-rate}
\end{eqnarray}
This latest form, which is symmetric under the interchange of neutrino and antineutrino momenta, is particularly convenient. It can be  further simplified by making use of the following identity:
\begin{eqnarray}
 \int  \frac{d^3\bm{p} d^3 \bm{q} }{q p} f(Q+P) Q^{\mu} P^{\nu}=\frac{\pi}{6}\int_{0}^{\infty} d\tilde{q}_0 \int d^3 \tilde{\bm{q}} \, 
 \theta(\tilde{q}_0^2-\tilde{q}^2) f(\tilde{Q})  \left( g^{\mu\nu} \tilde{Q}^2+2 \tilde{Q}^{\mu} \tilde{Q}^{\nu} \right) ,
 \label{integral-trick}
\end{eqnarray}
where $f(Q)$ is an arbitrary function. In Eq.~(\ref{integral-trick}), we used the following notation for the four-vectors:$Q=(q,\bm{q})$, $P=(p,\bm{p})$, and $\tilde{Q}\equiv (\tilde{q}_0,\tilde{\bm{q}})=Q+P$. The four-vector $\tilde{Q}$ represents the total energy ($\tilde{q}_0=p_0+q_0$) and momentum ($\tilde{\bm{q}}=\bm{p}+\bm{q}$) of the neutrino-antineutrino pair produced in a single weak process, $q_f \to q_f + \nu + \bar{\nu}$. Since both neutrinos and antineutrinos are massless here, their total four-momentum $\tilde{Q}$ is timelike, satisfying $\tilde{Q}^2=2q_0 p_0 (1-\cos\varphi_{\nu\bar{\nu}})\geq 0$, where $\cos\varphi_{\nu\bar{\nu}}$ denotes the angle between the momenta of the two particles.

For notational simplicity, we henceforth drop the tilde and denote the total four-momentum of the neutrino-antineutrino pair as $Q$. With this convention, we arrive at the following result:
\begin{eqnarray}
\dot{\cal E}_\nu &=&-  \frac{2 G_F^2}{3(2\pi)^5}
\int_{0}^{\infty} dq_0 \int_{q^2\leq q_0^2} d^3\bm{q} \, q_0 n_B( q_0) 
 \left( g^{\delta\sigma} Q^2 -  Q^{\delta} Q^{\sigma} \right) 
\Im \left[\Pi^R_{\delta\sigma}(Q)\right],
\label{E-dot-1}
\end{eqnarray}
where the timelike nature of $Q$ is enforced by limiting the integration range to $q^2\leq q_0^2$.

The imaginary part of the retarded self-energy, $\Im \left[\Pi^R_{\delta\sigma}(Q)\right]$, along with its Lorentz contraction with the lepton tensor, is calculated in Appendix~\ref{Z-self-energy}. Using the final result from Eq.~(\ref{res-L-ImPi}), we obtain the following expression for the energy emission rate:
\begin{eqnarray}
\dot{\cal E}_\nu &=& \frac{N_c G_F^2}{6 (2\pi)^6 \ell^2}  \sum_{n=0}^{\infty}  \sum_{s=1}^{\infty} 
\int_{-\infty}^{\infty} d k_z \int_{q^2\leq q_0^2}  d^3\bm{q} \,q_0
\frac{ n_F(E_{n^\prime,k_z+q_z}-\mu_{f}) \left[1-n_F(E_{n,k_z}-\mu_{f}) \right] }{E_{n^\prime,k_z+q_z} E_{n,k_z}} 
\nonumber \\
&\times& 
 \left[\left((c_V^f)^2+(c_A^f)^2\right) \tilde{S}_1 +\left((c_V^f)^2-(c_A^f)^2\right) \tilde{S}_2 + c_V^f c_A^f \tilde{S}_3\right] ,
\label{E-dot-30}
\end{eqnarray}
where $n_F(E-\mu)\equiv 1/\left[e^{(E-\mu)/T}+1\right]$ denotes the Fermi-Dirac distribution function, $E_{n,k_z} = \sqrt{2n |e_f B|+k_z^2+m^2}$ is the Landau-level energy, $\ell = 1/\sqrt{|e_f B|}$ is the magnetic field length, and $s=n^\prime-n\geq 1$ is the difference between the Landau-level indices of the initial and final quark states. The total energy of the neutrino and antineutrino pair is given by $q_0 \equiv E_{n^\prime,k_z+q_z} - E_{n,k_z}$. 

In deriving the final result in Eq.~(\ref{E-dot-30}), we also made use of the identity:
\begin{eqnarray}
n_B( q_0) \left[n_F(E_{n^\prime,k_z+q_z}-\mu_{f})-n_F(E_{n,k_z}-\mu_{f}) \right] 
= - n_F(E_{n^\prime,k_z+q_z}-\mu_{f}) \left[1-n_F(E_{n,k_z}-\mu_{f}) \right] .
\end{eqnarray}

The explicit forms of the functions $\tilde{S}_i$ are derived in Appendix~\ref{Z-self-energy}, with the final results given by:
\begin{eqnarray}
\tilde{S}_1&=& \left[-2 (q_0^2-q_z^2-q_\perp^2) \frac{n+n^\prime}{\ell^2} 
-m^2\left(q_0^2-q_z^2\right) \right]
\left(\mathcal{I}_{0}^{n,n^\prime}(\xi_q) +\mathcal{I}_{0}^{n-1,n^\prime-1}(\xi_q)\right) \nonumber\\
&+&\left[2 (q_0^2-q_z^2-q_\perp^2) \frac{n+n^\prime}{\ell^2} 
- \left(q_0^2-q_z^2 -q_\perp^2 \right)^2 
+m^2 \left(2q_0^2-2q_z^2-q_\perp^2\right)
\right] \left(\mathcal{I}_{0}^{n,n^\prime-1}(\xi_q) +\mathcal{I}_{0}^{n-1,n^\prime}(\xi_q)\right) , 
\label{S1-best}  \\
\tilde{S}_2&=&- m^2 \left(\mathcal{I}_{0}^{n,n^\prime}(\xi_q)+\mathcal{I}_{0}^{n-1,n^\prime-1}(\xi_q)\right) \left(q_0^2-q_z^2-2 q_\perp^2\right) 
- m^2 \left( \mathcal{I}_{0}^{n,n^\prime-1}(\xi_q)+\mathcal{I}_{0}^{n-1,n^\prime}(\xi_q)\right)  \left(2q_0^2-2q_z^2-q_\perp^2  \right)  ,   
\label{S2-best}  \\
\tilde{S}_3&=& 2 s_\perp  \left[ (k_z+q_z)E_{n} -k_z E_{n^\prime} \right] \left[\frac{2 (n^\prime-n)}{\ell^2} \left(\mathcal{I}_{0}^{n,n^\prime}(\xi_q)-\mathcal{I}_{0}^{n-1,n^\prime-1}(\xi_q)\right) + \left( \mathcal{I}_{0}^{n,n^\prime-1}(\xi_q)-\mathcal{I}_{0}^{n-1,n^\prime}(\xi_q)\right)  \left(2q_0^2-2q_z^2 - 3 q_\perp^2\right) 
 \right] , \nonumber\\
 \label{S3-best}
\end{eqnarray}
where $\xi_q =\bm{q}_\perp^2\ell^2/2$, and function $\mathcal{I}_{0}^{n,n^\prime}\left(\xi\right)$ is defined in terms of generalized Laguerre polynomials as follows \cite{Wang:2021ebh}:
\begin{eqnarray}
\mathcal{I}_{0}^{n,n^{\prime}}(\xi)&=& \frac{n!}{(n^\prime)!}e^{-\xi} \xi^{n^\prime-n} \left(L_{n}^{n^\prime-n}\left(\xi\right)\right)^2 .
\label{I0-nnprime}  
\end{eqnarray}
The final expression for the $\nu\bar{\nu}$ synchrotron emission rate in Eq.~(\ref{E-dot-30}) represents one of the main results of this study. Its general structure, along with the coefficient functions $\tilde{S}_i$, is in agreement with the results obtained via a different method in Ref.~\cite{Kaminker:1992su}.  It is worth noting, however, that by utilizing the definition of $q_0$ and the expressions for the Landau-level energies, we were able to present the function  $\tilde{S}_1$ in a more compact form.

\subsection{Approximate synchrotron emission rate for dense quark matter}
\label{sec:dense-quark-matter}

While the expression for the synchrotron emission rate in Eq.~(\ref{E-dot-30}) is exact, it is not particularly convenient for practical applications in dense quark matter under compact star conditions. The main challenges arise from the large hierarchy of energy scales: the quark chemical potentials (and, to a lesser extent, the electron chemical potential) are much larger than the temperature, magnetic field strength, and quark masses. In general, this implies that a large number of occupied Landau-level quark states participate in weak processes, and numerous quantum transitions between closely spaced states near the Fermi surface contribute to the synchrotron emission.

To address this hierarchy, it is essential to recognize that the relevant dynamics are governed by quark degrees of freedom near the Fermi surface. Indeed, states with energies far above the Fermi energy have exponentially small occupation numbers, while those well below the Fermi surface are Pauli blocked. As a result, only a narrow band of quark states, within an energy range of order $T$, can participate in weak processes.

Given that $\mu_{f} \gg T$, the only two relevant low-energy scales in the problem are the temperature $T$ and the Landau-level spacing at the Fermi surface, given by $\delta\epsilon_B \equiv |e_f B|/\mu_{f}$. (The quark mass $m$, which is also much smaller than $\mu_{f}$, does not play a significant role near the Fermi surface.) The ratio of these two low-energy scales, $b\equiv |e_f B|/(T\mu_{f})$, naturally defines two distinct regimes: the weak-field, high-temperature regime ($b \ll 1$), and the strong-field, low-temperature regime ($b \gg1$). 

In the weak-field (high-temperature) regime, a large number of transitions between closely spaced Landau levels (with $s_{\rm max}\gg 1/b$) contribute to the rate, effectively diminishing the impact of Landau-level quantization. Nevertheless, as $B\to 0$, the overall rate vanishes, since energy-momentum conservation becomes increasingly difficult to satisfy in this limit.
In contrast, in the strong-field (low-temperature) regime, the effect of Landau-level quantization becomes pronounced, and the rate is dominated by transitions between well-separated adjacent levels (i.e., $s\gtrsim 1$). As the Landau-level separation increases, the density of thermally excited quarks near the Fermi surface diminishes, leading to an exponential suppression of synchrotron radiation.

Taking into account that the Landau level separation at the Fermi surface $\delta\epsilon_B \equiv |e_f B|/\mu_{f}$ is much smaller than the quark chemical potential itself, the sum over index $n$ can be approximated by an integral, i.e.,
\begin{eqnarray}
\sum_{n=0}^{\infty}  f(2n|e_f B|) = \frac{\ell^2}{2\pi} \int d^2 \bm{k}_\perp f(k_\perp^2)  .
\end{eqnarray}
Before implementing this approximation, however, one requires a suitable approximation for the form factor function in Eq.~(\ref{I0-nnprime}). The corresponding large-$n$ behavior is given by 
\begin{eqnarray}
\mathcal{I}_{0}^{n,n+s}(\xi)& \simeq & \left[J_{s}\left(2\sqrt{\xi n}\right)\right]^2, 
\label{I00-approx}
\end{eqnarray}
where $s\ll n$ is assumed. In the derivation, we used Stirling's approximation for factorials and the large-$n$ asymptotes for the generalized Laguerre polynomials \cite{Szego:1975},
\begin{eqnarray}
L_{n}^{\alpha}(z) \simeq \left(\frac{n}{z}\right)^{\alpha/2} J_{\alpha}(2\sqrt{z n}). 
\end{eqnarray}
By replacing the sum over $n$ with an integral and applying the asymptotes in Eq.~(\ref{I00-approx}), the approximate expression for the synchrotron emission rate in dense quark matter takes the following form:  
\begin{eqnarray}
\dot{\cal E}_\nu &=& \frac{N_c G_F^2}{6 (2\pi)^6} \sum_{s=1}^{\infty} 
\int_0^{\infty}  k^2 dk \int_{0}^{\pi}  \sin\theta d\theta \int_{q^2\leq q_0^2}  d^3\bm{q} \,q_0
\frac{ n_F(E_{k}+q_0-\mu_{f}) \left[1-n_F(E_{k}-\mu_{f}) \right] }{E_{k} (E_{k}+q_0)} 
\nonumber \\
&\times& 
 \left[\left((c_V^f)^2+(c_A^f)^2\right) \tilde{S}_1 +\left((c_V^f)^2-(c_A^f)^2\right) \tilde{S}_2 + c_V^f c_A^f \tilde{S}_3\right] ,
\label{E-dot-40}
\end{eqnarray}
where $q_0 \simeq \left(k q_z \cos\theta+s |e_f B|\right)/E_k$, $E_{k}=\sqrt{k^2+m^2}$, and the functions $\tilde{S}_i$ are given by the same expressions as before, see Eqs.~(\ref{S1-best})--(\ref{S3-best}), except that the form factor functions $\mathcal{I}_{0}^{n,n+s}(\xi_q)$ are replaced by their asymptotic forms as in Eq.~(\ref{I00-approx}). Note also that $k_z =k \cos\theta$, $k_\perp =k \sin\theta$, and that the argument of the Bessel functions $2\sqrt{\xi_q n}$ reduces to $k_\perp q_\perp \ell^2$.

Due to the presence of Fermi-Dirac distribution functions in the integrand of Eq.~(\ref{E-dot-40}), the dominant contribution to the emission rate  comes from quark states with momenta $k$ in close vicinity of the Fermi surface. Therefore, the result can be well approximated by replacing the quark momentum with its Fermi surface value ($k\approx k_F\approx\mu_f$) throughout the integrand, except within the distribution functions themselves. Under this approximation, and up to higher-order corrections suppressed by inverse powers of the quark chemical potential, the integral over $k$ can be performed exactly, yielding the following result:
\begin{eqnarray}
\dot{\cal E}_\nu &=& \frac{N_c G_F^2\mu_{f}^2}{3  (2\pi)^5} \left((c_V^f)^2+(c_A^f)^2\right)  \sum_{s=1}^{\infty} 
\int_{0}^{\pi}\sin\theta d\theta \int_{-s \delta\epsilon_B/(1+\cos\theta)}^{s \delta\epsilon_B/(1-\cos\theta)} dq_z  
\int_{0}^{\sqrt{q_0^2-q_z^2}}  q_\perp dq_\perp \,q_0^2 \,  n_B(q_0) \nonumber\\
&&\times
 \left( \left[J_{s-1}\left(q_\perp\mu_{f} \ell^2 \sin\theta\right)\right]^2+\left[J_{s+1}\left(q_\perp\mu_{f} \ell^2 \sin\theta\right)\right]^2 -2\left[J_{s}\left(q_\perp\mu_{f} \ell^2 \sin\theta\right)\right]^2\right)  \left(q_0^2-q_z^2-q_\perp^2\right) \sin^2\theta ,
\label{E-dot-41}
\end{eqnarray}
where $q_0 \simeq q_z \cos\theta +s \delta\epsilon_B$. In deriving this expression, we employed the approximate forms of $\tilde{S}_i$ given in Eqs.~(\ref{S1-approx-app}) -- (\ref{S3-approx-app}) of Appendix~\ref{Z-self-energy}, which capture all leading-order contributions while neglecting subleading terms suppressed by inverse powers of the quark chemical potential.

Let us point out that the rate in Eq.~(\ref{E-dot-41}) contains an overall factor of $\mu_{f}^2$, which accounts for the degeneracy of quark states near the Fermi surface, whose area scales as $\mu_{f}^2$. After performing the integration over $q_\perp$ in Eq.~(\ref{E-dot-41}), we obtain
\begin{eqnarray}
\dot{\cal E}_\nu&=&\frac{2 N_c G_F^2 |e_fB|^2 }{3  (2\pi)^5} \left((c_V^f)^2+(c_A^f)^2\right)  \sum_{s=1}^{\infty} 
\int_{0}^{\pi} \sin\theta d\theta \int_{-s \delta\epsilon_B/(1+\cos\theta)}^{s \delta\epsilon_B/(1-\cos\theta)} dq_z  
 \,q_0^2 \,  n_B(q_0)(q_0^2-q_z^2) J_{s+1}\left(w\right)J_{s-1}\left(w\right),
\label{E-dot-41a}
\end{eqnarray}
where $w\equiv  \mu_{f} \ell^2 \sqrt{q_0^2-q_z^2} \sin\theta$. 

It is instructive to note that the explicit overall factor of $\mu_{f}^2$ disappears after the $q_\perp$ integration. The only remaining dependence on $\mu_{f}$ enters through the low-energy parameter $\delta\epsilon_B \equiv |e_f B|/\mu_{f}$. This implies that the total emission rate no longer scales with $\mu_{f}^2$, a result consistent with earlier findings reported in Refs.~\cite{Kaminker:1992su,Bezchastnov:1997ew}

Next, by changing the integration variable from $\theta$ to $y=\cos\theta$, and introducing the dimensionless variables $u_z \equiv q_z/T$ and $u_0 \equiv q_0/T = u_z  y +s b$, we finally derive
\begin{eqnarray}
\dot{\cal E}_\nu&=&\frac{2 N_c G_F^2|e_fB|^2 T^5}{3  (2\pi)^5} \left((c_V^f)^2+(c_A^f)^2\right) F(b),
\label{E-dot-final}
 \end{eqnarray}
where $b \equiv \delta\epsilon_B/T = |e_fB|/(T\mu_{f})$ is the dimensionless ratio of the two low-energy energy scales introduced earlier, 
and the scaling function $F\left(b\right) $ is defined by
\begin{eqnarray}
F\left(b\right) &=&\sum_{s=1}^{\infty} 
 \int_{-1}^{1}dy \int_{-sb/(1+y)}^{s b/(1-y)} d u_z \,
 \frac{\left(u_z y +s b\right)^2 \left[ \left(u_z y +s b\right)^2-u_z^2\right] }{e^{u_z y +s b}-1} J_{s+1}\left(\tilde{w}\right)J_{s-1}\left(\tilde{w}\right) ,
\label{F-b}
 \end{eqnarray}
with $\tilde{w}=\frac{1}{b} \sqrt{1-y^2} \sqrt{\left(y u_z +s b\right)^2-u_z^2}$.  

The expression in Eq.~(\ref{E-dot-final}) is the main analytical result used in the next section to compute numerically the $\nu\bar{\nu}$ synchrotron emission rate in dense quark matter. Depending on the value of the dimensionless parameter $b = |e_fB|/(T\mu_{f})$, two distinct regimes emerge in the limits of large and small $b$. 

As anticipated, in the regime $b\gg 1$, corresponding to strong magnetic fields or low temperatures, only the first few terms in the sum over $s$ contribute significantly. The resulting numerical computation is straightforward and yields an emission rate that decreases rapidly with increasing $b$, following an approximate scaling behavior dominated by $F(b) \sim \exp(-b/2)$.

In contrast, in the regime $b\ll 1$, realized in the limit of weak magnetic fields or high temperatures, a large number of transitions with a broad range of level spacings $s=n^\prime -n$ contribute to the rate. Since $F\left(b\right)$ approaches a finite, nonzero value as $b\to 0$, it follows that the emission rate (\ref{E-dot-final}) scales as $|e_fB|^2 T^5$ in this regime. Interestingly, as in the case of $\nu\bar{\nu}$ synchrotron emission from relativistic electrons \cite{Yakovlev:1981AN,Kaminker:1992su}, the dependence on the quark chemical potential $\mu_f$ is weak, entering only indirectly through the parameter $b$ inside $F\left(b\right)$. This is notable, as one might naively expect the rate to scale with the density of quark states on the Fermi surface, which grows as $\mu_f^2$. However, this phase-space enhancement is nearly completely compensated by the amplitude squared, which decreases as $1/\mu_f^2$ with increasing quark Fermi momentum. 

In our analysis of the synchrotron rate in Sec.~\ref{sec:results}, we rely on the numerical evaluation of the function $F\left(b\right)$ as defined in Eq.~(\ref{F-b}). The numerical calculation remains feasible even for relatively small values of $b\sim 0.01$. Nevertheless, it is useful to note that a simpler approximate analytic expression for $F\left(b\right)$ becomes available in the limit $b\to 0$. The corresponding approximation is discussed in the next subsection.

 \subsection{Synchrotron emission in the limit $ |e_fB|/(T\mu_{f}) \to 0$}
 \label{subsec:Synch-emission-b0}
 
In the small-$b$ limit, an approximate expression for the function $F(b)$, as defined in Eq.~(\ref{F-b}), can be obtained by replacing the discrete sum over $s$ with an integral over the continuous variable $v=sb$. However, since the dependence on $s$ appears within the order and arguments of the Bessel functions, it is convenient to employ the following asymptotic approximations \cite{Sokolov:1986}:
\begin{eqnarray}
 J_{s}\left( s x \right) &\simeq& \frac{\sqrt{1-x^2}}{\sqrt{3} \pi} K_{1/3}\left(\eta \right),
  \label{asympt-Js}\\
 J_{s}^\prime\left( s x\right)&\simeq& \frac{1-x^2}{\sqrt{3} \pi} K_{2/3}\left( \eta\right) ,
 \label{asympt-JsPrime}
\end{eqnarray}
with $\eta =\frac{s}{3}(1-x^2)^{3/2}$, which are valid for large $s$ and for $x$ approaching $1$ from below. These approximations are sufficient for evaluating the synchrotron emission rate, as the corresponding region provides the dominant contribution in the $b\to 0$ limit. The same approximations are used in deriving the usual electromagnetic synchrotron radiation in the quasiclassical limit \cite{Sokolov:1986}. 

By noting that 
\begin{equation}
 J_{s-1}\left( \tilde{w}\right)J_{s+1}\left(\tilde{w}\right) = \frac{s^2}{\tilde{w}^2}\left[J_{s}\left( \tilde{w}\right)\right]^2 -\left[J_{s}^\prime\left( \tilde{w}\right)\right]^2 ,
\end{equation}
and making use of the asymptotic forms of the Bessel functions given in Eqs.~(\ref{asympt-Js}) and (\ref{asympt-JsPrime}), we then derive the following result valid in the limit $b\to 0$:
\begin{equation}
 F(b)= \frac{1}{3\pi^2b}\int_{0}^{\infty} dv v^5 \int_{-1}^{1} dy \int_{-1/(1+y)}^{1/(1-y)} d u_z 
 \frac{\left(u_z y +1\right)^2 \left[ \left(u_z y +1\right)^2-u_z^2\right] }{e^{(u_z y +1)v}-1}
 \left( \frac{1-x^2}{x^2}  \left[K_{1/3}\left( \eta\right)\right]^2- (1-x^2)^2 \left[K_{2/3}\left( \eta\right)\right]^2  \right) ,
\end{equation}
where we rescaled the integration variable $u_z\to v u_z$. In the final expression, $x = \sqrt{1-y^2}\sqrt{\left(y u_z +1\right)^2-u_z^2}$ and $\eta =\frac{v}{3b}(1-x^2)^{3/2}$. As required for the validity of the Bessel function approximations used, the quantity $x$ is always less than $1$. This can be explicitly verified by noting that $1-x^2 = \left(y-u_z (1-y^2)\right)^2$ is positive definite.

As $b\to 0$, we find that $F\left(b\right)$ approaches a constant value of about $11.06$.

\section{Numerical results}
\label{sec:results}

In this section, using the analytical results from Eq.~(\ref{E-dot-final}), we perform a numerical study of the $\nu\bar{\nu}$ synchrotron emission rate in magnetized, dense quark matter under conditions relevant to compact stars. For simplicity, we assume that the matter consists of only the lightest quark flavors, i.e., up and down quarks. For the purposes of synchrotron emission, it is important to account for their distinct electric charges, given by $e_f = q_f e$ with $q_u = 2/3$ and $q_d = -1/3$, where $e$ is the absolute value of the electric charge of the electron. One should bear in mind that a small fraction of electrons must also be present to ensure charge neutrality and $\beta$-equilibrium of quark matter. 

In compact stars, quark matter is expected to reach densities around five times the nuclear saturation density. Under such conditions, the quark chemical potentials ($\mu_f$ with $f=u,d$) typically lie in the range of several hundred MeV. For our numerical calculations, we use representative values: $\mu_u = 300~\mbox{MeV}$, $\mu_d = 350~\mbox{MeV}$, and $\mu_e = 50~\mbox{MeV}$. As required, these values satisfy the $\beta$-equilibrium condition $\mu_d = \mu_u + \mu_e$. The temperature is unrestricted but should not be too high, e.g., $T \lesssim 10~\mbox{MeV}$, to avoid neutrino trapping. The background magnetic field is assumed to be strong but realistic for stars, ranging from $10^{14}~\mbox{G}$ to $10^{17}~\mbox{G}$.

In our analysis, we account for the production of all three neutrino flavors ($i = e, \mu, \tau$). Since synchrotron emission from quarks  proceeds only via neutral-current interactions, the total neutrino emission rate can be obtained by multiplying the single-flavor rate obtained in the preceding section by a factor of $N_{\nu} = 3$. This contrasts with synchrotron emission from electrons, which is mediated by both neutral- and charge-current interactions \cite{Yakovlev:1981AN,Kaminker:1992su}. When adding the synchrotron contribution from electrons in quark matter, we will add the same relevant neutral- and charge-current contributions as well. In the general expression for the rate in Eq.~(\ref{E-dot-final}), this amounts to using the appropriate coupling constants for each interaction channel, as specified in Table~\ref{cVcA-coupling}.

\subsection{Analysis of single-flavor rate}
\label{subsec:results-1flavor}

Before presenting the final numerical results for the total $\nu\bar{\nu}$ synchrotron emission rate in quark matter, we begin by examining the general features of a single-flavor contribution. First, we explore the role of different series of quantum transitions between quark Landau levels, characterized by the level separation $s = n^\prime- n$, where $n$ and $n^\prime$ are the initial and final Landau levels, respectively. The impact of these transitions on the emission rate is controlled by the dimensionless parameter $b = |e_f B| / (T \mu_f)$, which quantifies the relative importance of Landau-level quantization compared to the thermal energy scale. This dependence enters primarily through the function defined in Eq.~(\ref{F-b}), whose partial contributions from transitions of a given level spacing $s$ are denoted by $F_s(b)$, such that the total contribution is given by the sum $F(b)=\sum_{s=1}^{\infty}F_s(b)$.

\begin{figure}[t]
\centering
  \subfigure[]{\includegraphics[width=0.45\textwidth]{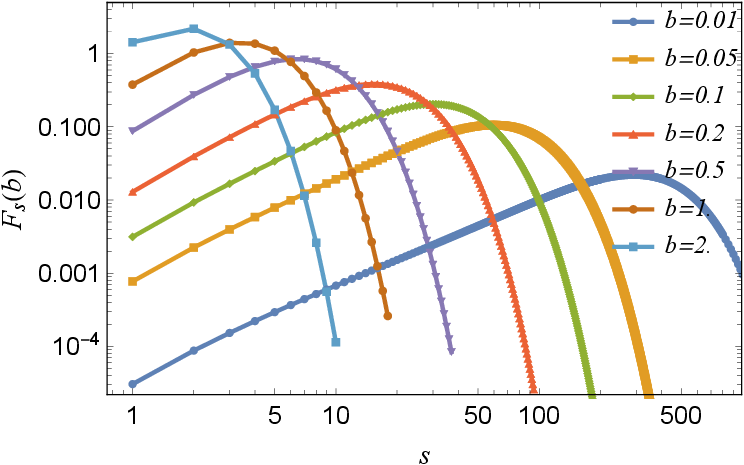}}
  \hspace{0.02\textwidth}
  \subfigure[]{\includegraphics[width=0.45\textwidth]{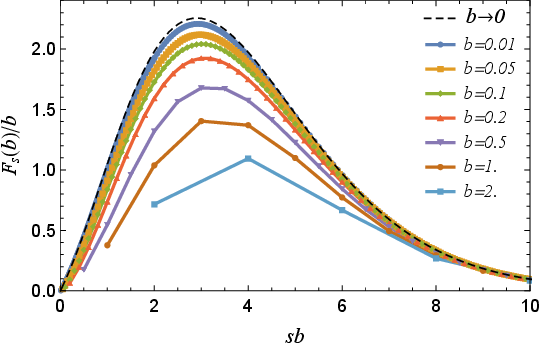}}
\caption{(a) Numerical data for partial contributions $F_s(b)$ to $F\left(b\right)$ defined in Eq.~(\ref{F-b}) as a function of level spacing $s=n^\prime -n$ for several fixed values of $b= |e_f B| / (T \mu_f)$,  (b) Ratio $F_s(b)/b$ as a function of the product $sb$.}
\label{fig:F-vs-s}
\end{figure}

Figure~\ref{fig:F-vs-s}(a) shows numerical results for $F_s(b)$ as a function of $s$ at several fixed values of $b$. For small values of $b$, where the Landau-level spacing is small compared to the thermal energy scale, transitions across a wide span of $s$ contribute significantly. In this regime, the dominant contributions come from around a peak located at $s_{\rm max} \simeq 3/b$, and the maximum value $F_{s_{\rm max}}(b)$ scales approximately as the first power of $b$. To make this scaling behavior more transparent, we plot the ratio $F_s(b)/b$ as a function of the product $sb$, as shown in Fig.~\ref{fig:F-vs-s}(b). In the limit $b \to 0$, the function approaches a fixed profile, shown by the dashed line.

%%%%%%%%%% FIGURE 1 %%%%%%%%%% 
%%%%%%%%%%%%%%%%%%%%

As the value of $b$ increases, the number of contributing transitions decreases, and only those with smaller level-separation values $s$ remain relevant. Starting from $b \gtrsim 2.6$, the transitions between adjacent Landau levels ($s = 1$) give the largest contribution, although transitions with a few other values of $s$ greater than $1$ still contribute substantially. For even stronger fields, i.e., $b \gtrsim 8$, transitions with $s \geq 2$ become nearly negligible, and synchrotron emission arises almost entirely from $s=1$ transitions. However, due to the large energy gap between Landau levels in this regime, the total emission rate becomes exponentially suppressed.

To gain deeper insight into the quark phase space responsible for synchrotron emission, it is instructive to examine the integrand in the definition of $F_s(b)$ in Eq.~(\ref{F-b}), prior to performing the final integration over the angular variable $y = \cos\theta$. Recall that $\theta$ denotes the angle between the quark momentum on the Fermi surface and the direction of the magnetic field. The limiting values $y = \pm 1$ correspond to the poles of the Fermi surface, while $y = 0$ represents the equatorial plane perpendicular to the field.

\begin{figure}[t]
\centering
  \subfigure[]{\includegraphics[width=0.32\textwidth]{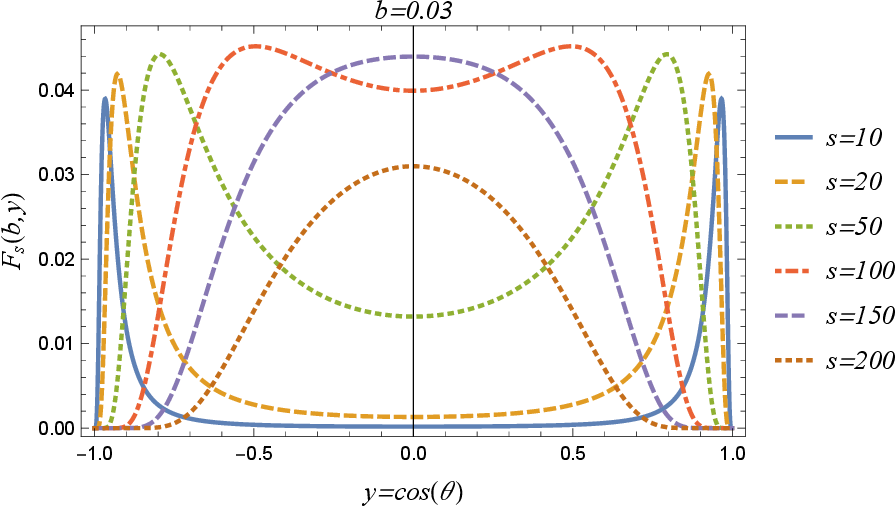}}
\hspace{0.0\textwidth}
 \subfigure[]{\includegraphics[width=0.32\textwidth]{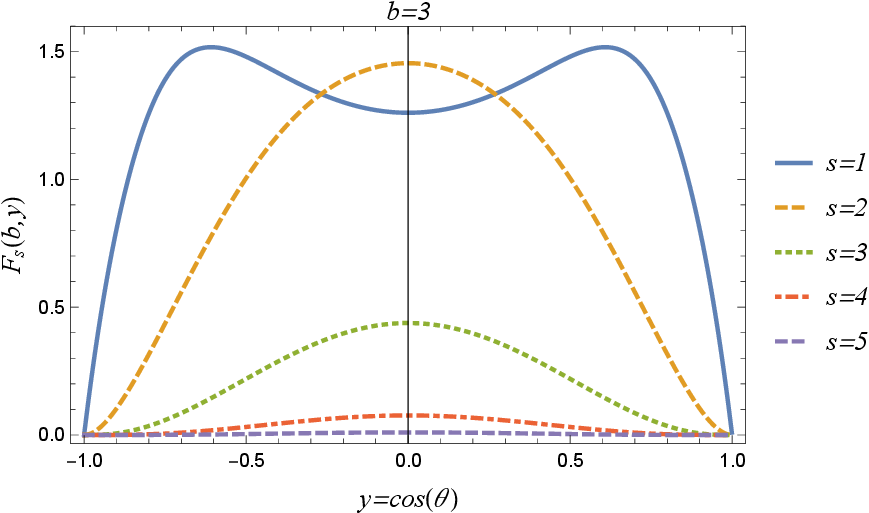}}
\hspace{0.0\textwidth}
 \subfigure[]{\includegraphics[width=0.32\textwidth]{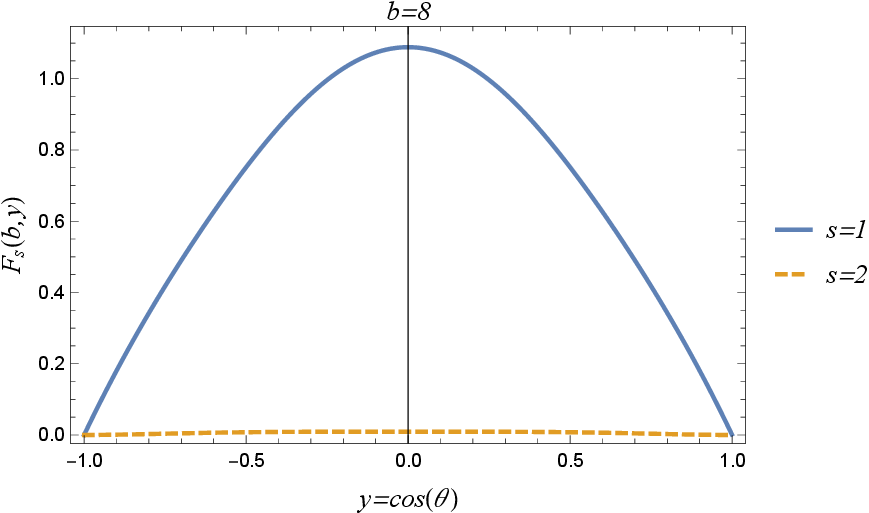}}
\caption{(a) Representative numerical results for $F_s(b,y)$ as a function  of $y=\cos\theta$ at fixed $b=0.03$ and several values of $s$. (b), (c) $F_s(b,y)$ for two larger values of $b$, i.e., $b=3$ and $b=8$, respectively, including results for the lowest few $s$ values in each case.}
\label{fig:Fsb-cosTheta}
\end{figure}

Representative plots of the integrand, denoted by $F_s(b,y)$, as a function of $y = \cos\theta$ are presented in Fig.~\ref{fig:Fsb-cosTheta}, with each panel corresponding to a different value of the dimensionless parameter $b$. Within each panel, multiple curves are shown for selected fixed values of the Landau-level separation $s$. These visualizations illustrate how the angular distribution of synchrotron-emitting quark states varies with both $b$ and $s$, offering valuable intuition about the underlying emission dynamics.

%%%%%%%%%% FIGURE 1 %%%%%%%%%% 
%%%%%%%%%%%%%%%%%%%%

For small values of $b$, the dominant regions on the Fermi surface contributing to synchrotron emission vary significantly with the value of $s$, see Fig.~\ref{fig:Fsb-cosTheta}(a). When $s$ is relatively small ($s \ll s_{\rm max}\simeq 3/b$), the dominant contributions come from regions with large $|y|$, corresponding to quark momenta that lie relatively far from the equatorial plane ($y=0$), but not too close to the poles of the Fermi sphere. As $s$ increases and approaches the optimal value $s_{\rm max} \simeq 3/b$, the dominant contributions shift toward a broad band of quark states near the equator, typically within $|y| \lesssim 0.5$. For even larger $s$ ($s \gg s_{\rm max}$), the width of the relevant band narrows down, indicating that fewer states near $y = 0$ contribute efficiently to the synchrotron rate.

In contrast, the case of large $b$, illustrated by Fig.~\ref{fig:Fsb-cosTheta}(b) and (c), is significantly simpler. Here, synchrotron emission is dominated by transitions between adjacent Landau levels ($s = 1$), and the quark states near the equatorial plane ($y=0$) account for the majority of the emission. As seen from Fig.~\ref{fig:Fsb-cosTheta}(c), this feature becomes increasingly pronounced with larger $b$, even though the total emission rate is strongly suppressed due to the quantization of Landau levels.

By summing over all relevant Landau-level transitions, we obtain function $F(b)$, with results shown as linear and logarithm plots in panels (a) and (b) of Fig.~\ref{fig:F-vs-b}, respectively. (Numerical data are provided in Supplemental Material \cite{DataF:2025}.) As seen from the plots, $F(b)$ is a monotonically decreasing function of the low-energy parameter $b$. It starts from a finite value at $b = 0$, specifically $F(0)\approx 11.06$, and falls off rapidly as $b$ increases. This behavior implies that, in the low-$b$ limit, the single-flavor synchrotron emission rate approaches the following asymptotic behavior:
\begin{eqnarray}
\dot{\cal E}_\nu& \simeq &  \frac{2 N_c G_F^2|e_fB|^2 T^5}{3 (2\pi)^5} \left((c_V^f)^2+(c_A^f)^2\right) F(0) , 
\qquad \mbox{for}\qquad 
T\gg \frac{|e_f B|}{\mu_{f}} .
 \end{eqnarray}
 As expected, the synchrotron rate vanishes in the $B\to 0$ limit.
 
In the opposite regime, where $b$ becomes large, the function is exponentially suppressed: $F\left(b\right) \propto  \exp(-b/2) $. This behavior reflects the strongly reduced thermal activation of quark states quantized in Landau levels near the Fermi surface.
Remarkably, over a broad range of $b$ values, i.e., from $b \approx 0$ up to $b \approx 100$, the numerical dependence of $F\left(b\right) $  is accurately captured by the following analytical fit: 
\begin{eqnarray}
F\left(b\right) &\approx &\frac{11.06 - 0.46 b  + 1.13 b^2 + 0.017 b^3 + 0.00165 b^4}{1 - 0.022 b + 1.4\times 10^{-4}  b^3 + 7.7 \times 10^{-5} b^4 + 1.5\times 10^{-7} b^5}e^{-b/2} .
 \end{eqnarray}
This expression reproduces the full numerical results to within approximately $3\%$, providing a convenient and accurate analytical form for estimating the synchrotron emission rate in quark matter. It may also prove useful in modeling synchrotron emission in other relativistic systems.

%%%%%%%%%% FIGURE 1 %%%%%%%%%% 
\begin{figure}[t]
\centering
  \subfigure[]{\includegraphics[width=0.45\textwidth]{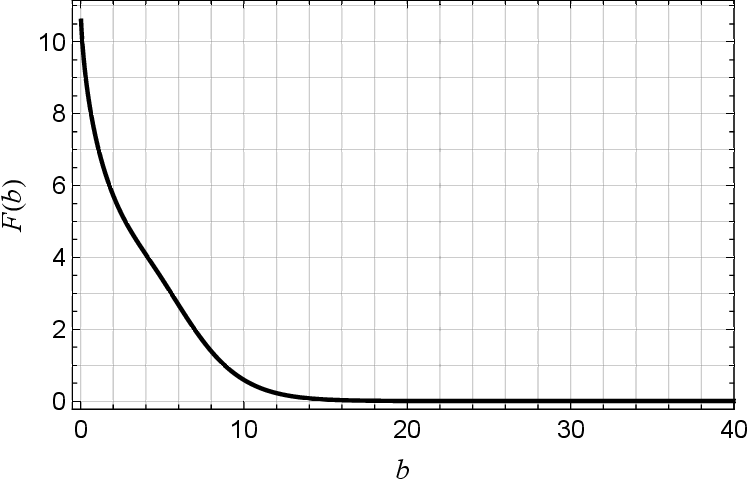}}
\hspace{0.02\textwidth}
  \subfigure[]{\includegraphics[width=0.45\textwidth]{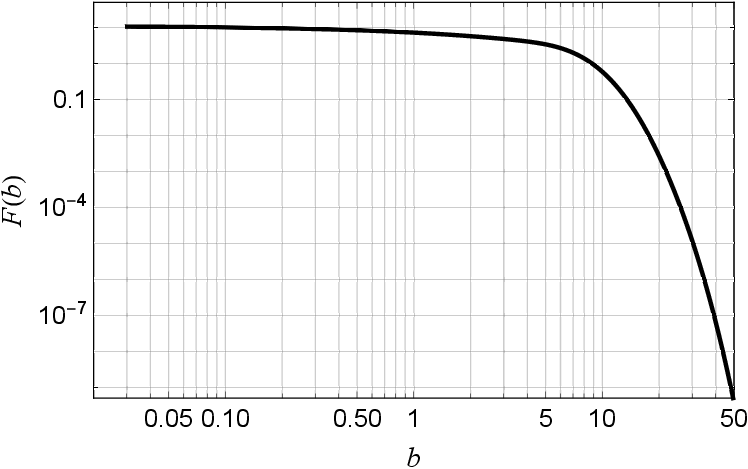}}
\caption{Linear (a) and logarithmic (b) plots of the numerical results for the function $F\left(b\right)$, defined in Eq.~(\ref{F-b}), which determines the single-flavor synchrotron neutrino emission rate.}
\label{fig:F-vs-b}
\end{figure}
%%%%%%%%%%%%%%%%%%%%

\subsection{Synchrotron emission rate in two-flavor quark matter}
\label{subsec:results-quark-matter}

Let us now turn to synchrotron emission from two-flavor dense quark matter. By adapting the general expression for the emission rate given in Eq.~(\ref{E-dot-final}) to include the contributions from both up and down quarks, we obtain
\begin{eqnarray}
\dot{\cal E}^{\rm (quarks)}_\nu&=& \sum_{f=u,d}\frac{2 N_c N_{\nu}G_F^2|e_fB|^2 T^5}{3  (2\pi)^5} \left((c_V^f)^2+(c_A^f)^2\right) F\left(\frac{|e_fB|}{\mu_f T}\right),
\label{E-dot-quarks}
 \end{eqnarray}
where $N_{\nu}=3$ accounts for the emission of all three neutrino flavors ($i = e, \mu, \tau$). 

As mentioned earlier, this quark contribution must be supplemented by the emission from electrons, yielding \cite{Kaminker:1992su}
\begin{eqnarray}
\dot{\cal E}^{\rm (electrons)}_\nu&=&\frac{2 G_F^2|eB|^2 T^5}{3  (2\pi)^5} \left[\left((c_V^{\rm eff})^2+(c_A^{\rm eff})^2\right)
+(N_{\nu}-1)\left( (c_V^e)^2+(c_A^e)^2\right) \right] F\left(\frac{|eB|}{\mu_eT}\right),
\label{E-dot-electrons}
 \end{eqnarray}
where the effective coupling constants $c_V^{\rm eff}$  and $c_A^{\rm eff}$ incorporate both neutral- and charged-current interactions relevant for $\nu_{e}\bar{\nu}_{e}$ emission  \cite{Yakovlev:1981AN,Kaminker:1992su}. In contrast, the constants $c_V^{e}$  and $c_A^{e}$ represent the standard neutral-current couplings responsible for emission of the remaining $N_{\nu}-1$ neutrino flavors.
 
\begin{figure}[t]
\centering
\subfigure[]{\includegraphics[width=0.45\textwidth]{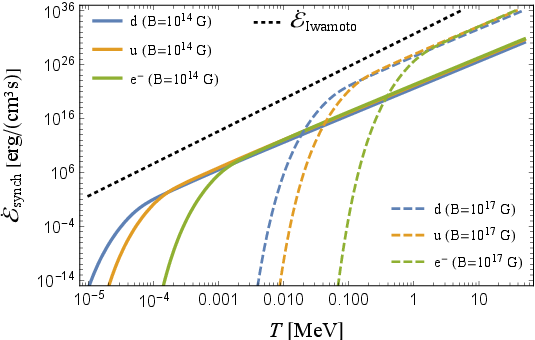}}
\hspace{0.02\textwidth}
\subfigure[]{\includegraphics[width=0.45\textwidth]{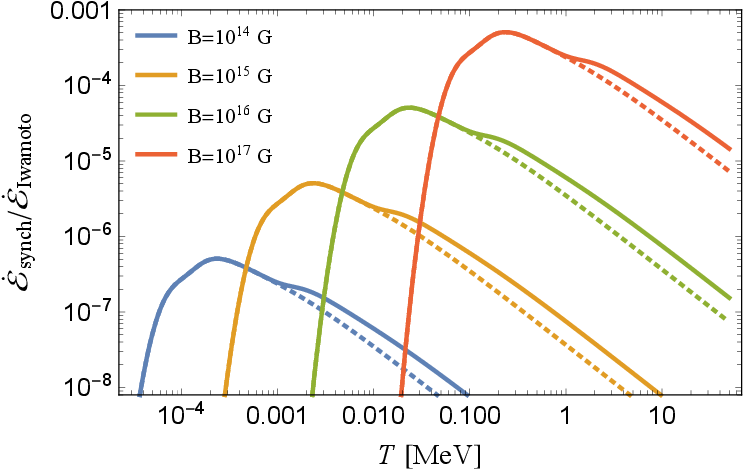}}
\caption{(a) Temperature dependence of the synchrotron neutrino emission rates from quarks and electrons in two-flavor dense quark matter for two magnetic field strengths: $B=10^{14}~\mbox{G}$ (solid lines) and $B=10^{17}~\mbox{G}$ (dashed lines). For comparison, Iwamoto's direct Urca emission rate is also plotted (black dotted line). 
(b) Ratio of the total synchrotron neutrino emission rate to the direct Urca emission rate as a function of temperature. Dotted lines show the contribution from quarks alone, while solid lines include both quark and electron contributions.}
\label{fig:rates-vs-T}
\end{figure}

The temperature dependence of synchrotron emission for each type of charged fermion is presented in Fig.~\ref{fig:rates-vs-T}(a). To highlight the role of the magnetic field, numerical results are shown for two field strengths: $B=10^{14}~\mbox{G}$ (solid lines) and $B=10^{17}~\mbox{G}$ (dashed lines). For comparison, we also plot Iwamoto's direct Urca emission rate (black dotted line), defined by \cite{Iwamoto:1980eb,Iwamoto:1982zz}:
\begin{equation}
 \dot{\cal E}_{\rm Iwamoto} \simeq\frac{457}{630} \alpha_s G_F^2\cos^2\theta_C \mu_u \mu_d \mu_e T^6 +O\left(\alpha_s^2,\frac{\mu_e}{\mu_u}\right),
  \label{E-dot-Iwamoto}
  \end{equation}
where $\theta_C$ is the Cabibbo angle, with $\cos^2\theta_C \approx 0.948$. For the strong coupling constant, we use the value $\alpha_s=0.3$.

We note in passing that the effect of a strong magnetic field on the direct Urca rate in quark matter was studied in detail in our earlier work \cite{Ghosh:2025sjn}. It was shown there that magnetic fields up to $10^{17}~\mbox{G}$ modify the average energy emission rate by only about $20\%$. For this reason, and to provide a clear benchmark, we chose to compare the synchrotron rates to Iwamoto's field-free Urca rate in Fig.~\ref{fig:rates-vs-T}.

%%%%%%%%%% FIGURE 1 %%%%%%%%%% 
%%%%%%%%%%%%%%%%%%%%

As the figure illustrates, synchrotron emission rates for both quarks and electrons are suppressed by several orders of magnitude relative to the direct Urca rate, even at the extreme field strength of $B=10^{17}~\mbox{G}$. This suppression is shown more clearly in Fig.~\ref{fig:rates-vs-T}(b), where we plot the ratio of the total synchrotron emission rate to the zero-field direct Urca rate (\ref{E-dot-Iwamoto}) as a function of temperature for several fixed values of the magnetic field strength. In this figure, dotted lines represent the contributions from quarks alone, while solid lines include both quark and electron contributions. Notably, electron synchrotron emission becomes significant at higher temperatures, although the total rate still remains much smaller than that of the direct Urca process across the temperature range considered.

At first glance, the $T^5$ temperature scaling of the synchrotron emission rate might suggest that it could dominate over the direct Urca rate, which scales as $T^6$, particularly at lower temperatures. Indeed, as seen from the high-temperature behavior of the rates in Fig.~\ref{fig:rates-vs-T}(a), the slope of the synchrotron curve is indeed smaller. However, this advantage gets lost as the temperature drops below the Landau-level energy separation and the synchrotron rate begins to decrease exponentially, preventing it from overtaking the direct Urca rate. A closer examination of Eqs.~(\ref{E-dot-quarks}) and (\ref{E-dot-Iwamoto}) further reveals that the synchrotron rate remains much smaller not only due to its specific temperature dependence, but also because, unlike the Urca rate, it does not scale with the quark density of states at the Fermi surface. In addition, the synchrotron rate has a significantly smaller overall prefactor, $(2\pi)^{-5} \simeq 10^{-4}$, which has no analogue in the direct Urca rate, further contributing to its relative suppression.

Let us emphasize that the approximations employed in our calculation are strictly valid in the regime where the temperature is much smaller than the quark chemical potential. In Fig.~\ref{fig:rates-vs-T}, however, we extend the results to temperatures up to $50~\mbox{MeV}$, which approach or may slightly exceed the regime of strict validity. Therefore, the results at the highest temperatures should be interpreted with caution and regarded as extrapolations rather than quantitatively precise predictions.

\section{Summary}
\label{sec:Summary}

In this work, we developed a systematic approach to compute the synchrotron neutrino emission rate from magnetized dense quark matter. One of the distinguished features of our study is the use of the Kadanoff-Baym formalism, which allows for a more streamlined derivation by expressing the emission rate in terms of Green functions rather than explicit particle wave functions. This method offers practical advantages for studying many-body systems in strong magnetic fields.

We carried out a detailed analysis of the phase space relevant for synchrotron emission and identified the dominant regions on the quark Fermi surface that contribute to the rate. Our results show that the quark states near the equator of the Fermi surface (i.e., those with momenta nearly perpendicular to the magnetic field) play the leading role in synchrotron emission.

Furthermore, in relativistic dense quark matter relevant for compact stars, we demonstrated that the emission rate is governed by a single dimensionless parameter, namely the ratio of the Landau level spacing at the Fermi surface, $|e_f B|/\mu_{f}$, to the temperature $T$. When $|e_f B|/(\mu_{f} T)\ll 1$, many closely spaced Landau levels contribute collectively to the emission. In the opposite regime, $|e_f B|/(\mu_{f} T)\gg 1$, the rate is dominated by transitions between adjacent Landau levels. While the emission rate grows with the magnetic fields, our results show that even at field strengths as high as $B=10^{17}~\mbox{G}$, which are likely the strongest fields expected in compact stars, the synchrotron neutrino emission remains suppressed by more than 3 orders of magnitude compared to the direct Urca process. 

Generally, $\nu\bar{\nu}$ synchrotron emission in dense quark matter is similar to its counterpart in white dwarfs and neutron stars, where the process is dominated by electrons \cite{Yakovlev:1981AN,Kaminker:1992su,Kaminker:1993ey,Vidaurre:1995iv,Bezchastnov:1997ew}. In both cases, the synchrotron rate exhibits a power-law dependence on temperature, which suggests that it can compete with or even surpass the direct Urca process under certain conditions. However, the actual situation in quark matter differs in important ways. In white dwarfs and neutron stars, synchrotron emission typically competes with the modified Urca process, which has a steeper temperature dependence and is often subdominant at low temperatures. In contrast, in quark matter, the direct Urca process is kinematically allowed and remains a dominant channel for neutrino emission across a wide range of temperatures. One might naively expect that the large quark chemical potentials would mitigate the effects of Landau-level quantization, delaying the onset of exponential suppression in synchrotron emission to much lower temperatures and thus enhancing its competitiveness. However, detailed analysis shows that synchrotron emission never dominates the neutrino emissivity in any realistic regime of dense quark matter relevant to compact stars.

In summary, our study provides a detailed analysis of synchrotron neutrino emission in quark matter and demonstrates its limited impact on the cooling of magnetized compact stars with quark matter cores. In this work, we have focused exclusively on unpaired quark matter. However, at the densities typical of compact star interiors, quark matter may exist in a color-superconducting phase \cite{Shovkovy:2004me,Alford:2007xm}, which could significantly modify the emission processes and warrants further investigation. In the two-flavor color-superconducting phase, where some quark quasiparticles remain ungapped, synchrotron emission is expected to resemble that in unpaired matter. In contrast, in the color-flavor-locked phase \cite{Alford:1998mk}, where all quark quasiparticles are gapped, both direct Urca and synchrotron emission are strongly suppressed, suggesting that cooling proceeds through mechanisms unrelated to neutrino emission \cite{Shovkovy:2002kv}. More subtle behavior may occur in spin-$1$ color-superconducting phases \cite{Schmitt:2004et}, where gapless nodes or lines on the Fermi surface could significantly modify the relative efficiency of synchrotron and Urca processes \cite{Schmitt:2005wg}. A detailed investigation of these effects is left for future work.

\begin{acknowledgments}
This research was funded in part by the U.S. National Science Foundation under Grant No.~PHY-2209470.
\end{acknowledgments}

\section*{Data Availability}

The data that support the findings of this article are openly available \cite{DataF:2025}.

\appendix

\section{Vector $Z$-boson self-energy and its Lorentz contraction with the lepton tensor}
\label{Z-self-energy}

As we show in the main text, the expression for the neutrino synchrotron emission rate is given in terms of the imaginary part of the retarded vector boson self-energy  $\Pi^R_{\delta\sigma}(Q)$, see Eq.~(\ref{E-dot-1}).

The retarded self-energy $\Pi^R_{\delta\sigma}(Q)$ is defined by the following one-loop expression:
\begin{eqnarray}
 \Pi^{R}_{\delta\sigma}(Q) &=& -  i N_c   \sumint \frac{d^4K}{(2\pi)^4}\sum_{f=u,d}\mbox{Tr}\left[\gamma_\delta(c_V^q-c_A^q\gamma_5) \bar{S}_{f}(K)\gamma_\sigma(c_V^q-c_A^q\gamma_5) \bar{S}_{f}(K+Q)\right] ,
 \label{Pi-tensor}
\end{eqnarray}
where we use the following shorthand notation for the momentum integration and the sum over the Matsubara frequencies $\omega_k=(2k+1)\pi T$:
\begin{equation}
\sumint \frac{d^4 K}{(2\pi)^4} f(k_0,\bm{k}) = T\sum_{k=-\infty}^{\infty}  i  \int \frac{d^3 \bm{k}}{(2\pi)^3} f(i\omega_k,\bm{k}).
\label{int-Matsubara}
\end{equation}
The Fourier transform of the translationally invariant part of the quark propagator in the Landau-level representation is given by \cite{Miransky:2015ava} 
\begin{eqnarray}
\bar{S}_{f} (K)&=& i e^{-k_\perp^2\ell^2} \sum_{n=0}^{\infty} \sum_{\lambda=\pm} \frac{(-1)^n  D_{n}(K) }{E_{n,k_z}
\left[k_0+\mu_{f}+i\epsilon \, \sign(k_0)-\lambda E_{n,k_z}\right]} ,  
\label{prop-quark}
 \end{eqnarray} 
 where
 \begin{eqnarray}
 D_{n}(K)&=&\left[E_{n,k_z} \gamma^0+\lambda(m-k_{z}\gamma^3)\right] 
 \left[{\cal P}_{+}L_n\left(2 k_\perp^2\ell^2\right)
-{\cal P}_{-}L_{n-1}\left(2 k_\perp^2\ell^2\right)\right]
+2\lambda (\bm{k}_\perp\cdot\bm{\gamma}_\perp) L_{n-1}^1\left(2 k_\perp^2 \ell^2\right) .
 \label{Dn-quark}
\end{eqnarray}
Here $ E_{n,k_z}=\sqrt{2n|e_f B|+k_{z}^2+ m^2 }$ are the quark Landau-level energies, $\ell =1/\sqrt{|e_fB|}$ is the magnetic length, ${\cal P}_{\pm}=(1\pm i s_\perp \gamma^1\gamma^2)/2$ are the spin projectors, $s_\perp = \mbox{sign}(e_f B)$, and $L_n^{\alpha}\left(z\right)$ are the generalized Laguerre polynomials.

By substituting the explicit form of the quark propagator in Eq.~(\ref{prop-quark}) into Eq.~(\ref{Pi-tensor}), calculating the Matsubara sum, and performing the analytical continuation $q_0\to q_0+ i \epsilon$, we derive the following expression for the imaginary part of the self-energy:
\begin{eqnarray}
\Im \left[\Pi^{R}_{\delta\sigma}(Q)\right] &=& -\frac{N_c}{8 \pi^2} \sum_{n,n^\prime=0}^{\infty} (-1)^{n+n^\prime}  \int  d^3 \bm{k} 
\frac{ e^{-\bm{k}_\perp^2\ell^2-(\bm{k}_\perp+\bm{q}_\perp)^2\ell^2} }
 {E_{n,k_z} E_{n^\prime,k_z+q_z}} 
 \left[n_F(E_{n^\prime,k_z+q_z}-\mu_{f})-n_F(E_{n,k_z}-\mu_{f})\right] \nonumber\\
&\times&  
 T_{\delta\sigma}(K,K+Q) \delta\left(E_{n^\prime,k_z+q_z}-E_{n,k_z}-q_0\right) ,
 \label{Im_Pi-app}
 \end{eqnarray} 
 where
 \begin{eqnarray}
 T_{\delta\sigma}(K,K+Q)&=& \mbox{Tr}\left[\gamma_\delta(c_V^q-c_A^q\gamma_5) D_{n}(K)\gamma_\sigma(c_V^q-c_A^q\gamma_5) D_{n^\prime}(K+Q)\right].
\end{eqnarray}
Note that in dense quark matter, only quark states near the Fermi surface contribute significantly to synchrotron emission. This is because the thermal population of antiquarks is strongly suppressed at temperatures much lower than the quark chemical potential ($T\ll \mu_f$). Mathematically, this suppression arises from the form of the distribution functions (with a large positive value of $\mu_f$) and the energy-conserving $\delta$-function, which together strongly favor quark over antiquark contributions. Therefore, in the last expression, we kept only the contributions of quarks ($\lambda=1$) and neglected the contributions of antiquarks ($\lambda=-1$). 

In the neutrino emission rate in Eq.~(\ref{E-dot-1}), the imaginary part of the self-energy is contracted with another tensor, i.e., $g^{\delta\sigma} Q^2 -  Q^{\delta} Q^{\sigma}$. It is convenient, therefore, to first calculate for the corresponding Lorentz contraction with the tensor inside the integrand of Eq.~(\ref{Im_Pi-app}). The result reads
\begin{eqnarray}
\left( g^{\delta\sigma} Q^2 -  Q^{\delta} Q^{\sigma} \right)  T_{\delta\sigma}(K,K+Q)
=\left((c_V^f)^2+(c_A^f)^2\right) S_1 +\left((c_V^f)^2-(c_A^f)^2\right) S_2 + c_V^f c_A^f S_3 .
\end{eqnarray}
The explicit forms of functions $S_{i}$ are given by
\begin{eqnarray}
S_1&=&-2 (L_{n}L_{n^\prime}+L_{n-1} L_{n^\prime-1}) \left(E_{n} E_{n^\prime} -k_z (k_z+q_z)\right) \left(q_0^2-q_z^2\right)
-4(L_{n}L_{n^\prime}+L_{n-1} L_{n^\prime-1}) \left(q_0 k_z- q_zE_{n} \right) \left(q_0 (k_z+q_z)- q_zE_{n^\prime} \right) \nonumber \\
&&+2 (L_{n}L_{n^\prime-1}+L_{n-1} L_{n^\prime}) \left(E_{n} E_{n^\prime} -k_z (k_z+q_z)\right) \left(2q_0^2-2q_z^2-q_\perp^2\right)
\nonumber \\
&&-8L_{n-1}^{1} (L_{n^\prime}-L_{n^\prime-1}) \left(q_0 E_{n^\prime} - q_z  (k_z+q_z) \right) (\bm{q}_\perp \cdot  \bm{k}_\perp)
 -8 (L_{n}-L_{n-1})  L^{1}_{n^\prime-1} \left(q_0 E_{n} - q_z  k_z \right) (\bm{q}_\perp \cdot (\bm{k}_\perp+\bm{q}_\perp))\nonumber\\
&&+16L_{n-1}^{1}L^{1}_{n^\prime-1}\left(q_0^2-q_z^2-q_\perp^2\right)(\bm{k}_\perp \cdot (\bm{k}_\perp+\bm{q}_\perp))
-32L_{n-1}^{1}L^{1}_{n^\prime-1} (\bm{q}_\perp \cdot  \bm{k}_\perp) (\bm{q}_\perp \cdot (\bm{k}_\perp+\bm{q}_\perp))
, \\
S_2&=&2 m^2(L_{n}L_{n^\prime}+L_{n-1}L_{n^\prime-1})\left(q_0^2-q_z^2-2 q_\perp^2\right) 
-2 m^2 (L_{n}L_{n^\prime-1}+L_{n-1}L_{n^\prime})\left(2q_0^2-2q_z^2-q_\perp^2  \right)  , \\
S_3&=&4 s_\perp  (L_{n}L_{n^\prime}-L_{n-1} L_{n^\prime-1}) 
\left(E_{n} (k_z+q_z) - E_{n^\prime}k_z \right) \left(q_0^2 -q_z^2\right)
+8 s_\perp  (L_{n}L_{n^\prime}-L_{n-1} L_{n^\prime-1}) 
\left( q_0 k_z- q_z E_{n} \right)  \left( q_0 E_{n^\prime}-q_z (k_z+q_z) \right)  \nonumber \\
&&+4 s_\perp  (L_{n}L_{n^\prime-1}-L_{n-1} L_{n^\prime}) 
\left(E_{n} (k_z+q_z)- E_{n^\prime}k_z \right) \left(2q_0^2-2q_z^2 -q_\perp^2\right)
\nonumber \\
&&+16 s_\perp (L_{n}+L_{n-1})  L^{1}_{n^\prime-1} \left(q_0 k_z -q_z E_{n} \right)(\bm{q}_\perp \cdot (\bm{k}_\perp+\bm{q}_\perp))
+16 s_\perp L_{n-1}^{1} (L_{n^\prime}+L_{n^\prime-1})  \left(q_0 (k_z+q_z) -q_z E_{n^\prime} \right)(\bm{q}_\perp \cdot \bm{k}_\perp) ,
\end{eqnarray}
where, for simplicity of presentation, we abbreviated the notations $E_{n} \equiv E_{n,k_z}$, $E_{n^\prime} \equiv E_{n^\prime,k_z+q_z}$, $L_{n}\equiv L_{n}\left(2\bm{k}_\perp^2\ell^2 \right)$, and $ L_{n^\prime} \equiv L_{n^\prime} \left[2(\bm{k}_\perp+\bm{q}_\perp)^2\ell^2 \right]$.

Therefore, for the Lorentz contraction of the  imaginary part of the self-energy with the lepton tensor, we obtain
\begin{eqnarray}
\left( g^{\delta\sigma} Q^2 -  Q^{\delta} Q^{\sigma} \right) \Im \left[\Pi^{R}_{\delta\sigma}(Q)\right] &=& -\frac{N_c}{8 \pi^2} \sum_{n,n^\prime=0}^{\infty} (-1)^{n+n^\prime}  \int  d^3 \bm{k} 
\frac{ e^{-\bm{k}_\perp^2\ell^2-(\bm{k}_\perp+\bm{q}_\perp)^2\ell^2} }
 {E_{n,k_z} E_{n^\prime,k_z+q_z}} 
 \left[n_F(E_{n^\prime,k_z+q_z}-\mu_{f})-n_F(E_{n,k_z}-\mu_{f})\right] \nonumber \\
&\times& \left[ \left((c_V^f)^2+(c_A^f)^2\right) S_1 +\left((c_V^f)^2-(c_A^f)^2\right) S_2 + c_V^f c_A^f S_3\right] 
\delta\left(E_{n^\prime,k_z+q_z}-E_{n,k_z}-q_0\right) .
\label{def-L-ImPi}
\end{eqnarray}
The integral over the transverse components of the quark momentum $\bm{k}_\perp$ can be easily performed using the following table integrals:
\begin{eqnarray}
\int d^2 \bm{k}_\perp  e^{-\bm{k}_\perp^2\ell^2-(\bm{k}_\perp+\bm{q}_\perp)^2\ell^2}
L_{n} \left(2\bm{k}_\perp^2\ell^2 \right) L_{n^\prime}  \left[2(\bm{k}_\perp+\bm{q}_\perp)^2\ell^2 \right]
 &=& \frac{\pi}{2\ell^2}  (-1)^{n+n^\prime} \mathcal{I}_{0}^{n,n^\prime}\left(\xi_q\right) \label{k-perp-int-0} ,\\
\int d^2 \bm{k}_\perp  e^{-\bm{k}_\perp^2\ell^2-(\bm{k}_\perp+\bm{q}_\perp)^2\ell^2}
\left[\bm{k}_\perp\cdot (\bm{k}_\perp+\bm{q}_\perp)\right]
L^{1}_{n} \left(2\bm{k}_\perp^2\ell^2 \right) L^{1}_{n^\prime}  \left[2(\bm{k}_\perp+\bm{q}_\perp)^2\ell^2 \right]  
&=& \frac{\pi}{8\ell^4}  (-1)^{n+n^\prime} \mathcal{I}_{2}^{n,n^\prime}\left(\xi_q\right) \label{k-perp-int-2} ,\\
\int d^2 \bm{k}_\perp  e^{-\bm{k}_\perp^2\ell^2-(\bm{k}_\perp+\bm{q}_\perp)^2\ell^2}
\left[(\bm{k}_\perp+\bm{q}_\perp)\cdot \bm{a}_\perp\right]
L_{n} \left(2\bm{k}_\perp^2\ell^2 \right) L^{1}_{n^\prime}  \left[2(\bm{k}_\perp+\bm{q}_\perp)^2\ell^2 \right]  
&=&\frac{\pi}{4\ell^3} (-1)^{n+n^\prime}  (\hat{\bm{q}}_\perp\cdot \bm{a}_\perp)\mathcal{I}_{1}^{n,n^\prime}\left(\xi_q\right) \label{k-perp-int-1a}  ,\\
\int d^2 \bm{k}_\perp  e^{-\bm{k}_\perp^2\ell^2-(\bm{k}_\perp+\bm{q}_\perp)^2\ell^2}
\left(\bm{k}_\perp\cdot \bm{a}_\perp\right)
L^{1}_{n} \left(2\bm{k}_\perp^2\ell^2 \right) L_{n^\prime}  \left[2(\bm{k}_\perp+\bm{q}_\perp)^2\ell^2 \right] 
&=&-\frac{\pi}{4\ell^3}(-1)^{n+n^\prime}  (\hat{\bm{q}}_\perp\cdot \bm{a}_\perp)\mathcal{I}_{1}^{n^\prime,n}\left(\xi_q\right) \label{k-perp-int-1b}  ,
\end{eqnarray}
and
\begin{eqnarray}
\int d^2 \bm{k}_\perp  e^{-\bm{k}_\perp^2\ell^2-(\bm{k}_\perp+\bm{q}_\perp)^2\ell^2}
\left(\bm{k}_\perp\cdot \bm{a}_\perp \right)
\left[(\bm{k}_\perp+\bm{q}_\perp)\cdot  \bm{b}_\perp \right]
L^{1}_{n} \left(2\bm{k}_\perp^2\ell^2 \right) L^{1}_{n^\prime}  \left[2(\bm{k}_\perp+\bm{q}_\perp)^2\ell^2 \right]  =  \frac{\pi}{16 \ell^4} (-1)^{n+n^\prime}  &&\nonumber\\
\times  \left[\left(\bm{a}_\perp\cdot \bm{b}_\perp \right) 
\mathcal{I}_{2}^{n,n^\prime}\left(\xi_q\right)+  \left[ (\bm{a}_\perp\cdot \bm{b}_\perp) 
-2(\hat{\bm{q}}_\perp\cdot \bm{a}_\perp) (\hat{\bm{q}}_\perp\cdot \bm{b}_\perp) \right]
\mathcal{I}_{3}^{n,n^\prime}\left(\xi_q\right) \right] &&,  \label{k-perp-int-23} 
\end{eqnarray}
where $\xi_q =\bm{q}_\perp^2\ell^2/2$ and functions $\mathcal{I}_{i}^{n,n^\prime}\left(\xi_q\right)$ are the same form factor as in Ref.~\cite{Wang:2021ebh}. In particular, 
\begin{eqnarray}
\mathcal{I}_{0}^{n,n^{\prime}}(\xi)&=&  \frac{n!}{(n^\prime)!}e^{-\xi} \xi^{n^\prime-n}
\left(L_{n}^{n^\prime-n}\left(\xi\right)\right)^2 ,
\label{I0f-LL-form1}  \\
\mathcal{I}_{1}^{n,n^{\prime}}(\xi)&=&  \sqrt{2\xi}  \frac{n!}{(n^\prime)!} e^{-\xi} \xi^{n^\prime-n} 
L_{n}^{n^\prime-n+1}\left(\xi\right) 
L_{n}^{n^\prime-n}\left(\xi\right) 
\label{I1f-LL-form1} , \\
\mathcal{I}_{2}^{n,n^{\prime}}(\xi)&=& 2 \frac{(n+1)!}{(n^\prime)!} e^{-\xi}  \xi^{n^\prime-n}  L_{n}^{n^\prime-n}\left(\xi\right)L_{n+1}^{n^\prime-n}\left(\xi\right)  , 
\label{I2f-LL-form1} \\
\mathcal{I}_{3}^{n,n^{\prime}}(\xi)&=&  - 2 \frac{(n+1)!}{(n^\prime)!} e^{-\xi} \xi^{n^\prime-n} 
L_{n}^{n^\prime-n+1}\left(\xi\right) 
L_{n+1}^{n^\prime-n-1}\left(\xi\right).
\label{I3f-LL-form1} 
\end{eqnarray}
After the integration over $\bm{k}_\perp$, we obtain the following result:
\begin{eqnarray}
\left( g^{\delta\sigma} Q^2 -  Q^{\delta} Q^{\sigma} \right) \Im \left[\Pi^{R}_{\delta\sigma}(Q)\right] &=& \frac{N_c}{8 \pi \ell^2} \sum_{n=0}^{\infty} \sum_{s=1}^{\infty}  \int  \frac{d k_{z}}{E_{n,k_z} E_{n^\prime,k_z+q_z}} 
 \left[n_F(E_{n^\prime,k_z+q_z}-\mu_{f})-n_F(E_{n,k_z}-\mu_{f})\right] \nonumber \\
&\times& \left[ \left((c_V^f)^2+(c_A^f)^2\right) \tilde{S}_1 +\left((c_V^f)^2-(c_A^f)^2\right) \tilde{S}_2 + c_V^f c_A^f \tilde{S}_3\right] 
\delta\left(E_{n^\prime,k_z+q_z}-E_{n,k_z}-q_0\right) ,
\label{res-L-ImPi}
\end{eqnarray}
where $n^\prime \equiv n+s$. It should be noted that the $\delta$-function enforces the total neutrino-antineutrino energy to match the energy difference between the initial and final quark states, i.e., $q_0=E_{n^\prime,k_z+q_z}-E_{n,k_z}$. Analyzing the corresponding phase space as in Ref.~\cite{Wang:2021ebh}, we find that $q_0>0$ requires $n^\prime >n$, so that the sum over $s\equiv n^\prime -n$ starts from $s=1$.

The new functions $\tilde{S}_i $ are defined by
\begin{eqnarray}
\tilde{S}_i &=& - \frac{\ell^2}{\pi} (-1)^{n+n^\prime} \int d^2 \bm{k}_\perp  e^{-\bm{k}_\perp^2\ell^2-(\bm{k}_\perp+\bm{q}_\perp)^2\ell^2} S_1 .\end{eqnarray}
Their explicit expressions read
\begin{eqnarray}
\tilde{S}_1&=& \left(\mathcal{I}_{0}^{n,n^\prime}+\mathcal{I}_{0}^{n-1,n^\prime-1}\right) \left(E_{n} E_{n^\prime} -k_z (k_z+q_z)\right) \left(q_0^2-q_z^2\right)
+2\left(\mathcal{I}_{0}^{n,n^\prime}+\mathcal{I}_{0}^{n-1,n^\prime-1}\right) \left(q_0 k_z- q_zE_{n} \right)^2 \nonumber \\
&&+ \left(\mathcal{I}_{0}^{n,n^\prime-1}+\mathcal{I}_{0}^{n-1,n^\prime}\right) \left(E_{n} E_{n^\prime} -k_z (k_z+q_z)\right) \left(2q_0^2-2q_z^2-q_\perp^2\right)  \nonumber \\
&&+\frac{2\sqrt{2\xi}}{\ell^2} \left(\mathcal{I}_{1}^{n^\prime,n-1}+\mathcal{I}_{1}^{n^\prime-1,n-1}\right) \left(q_0 E_{n^\prime} - q_z  (k_z+q_z) \right)  
-\frac{2\sqrt{2\xi}}{\ell^2}  \left(\mathcal{I}_{1}^{n,n^\prime-1}+\mathcal{I}_{1}^{n-1,n^\prime-1}\right) \left(q_0 E_{n} - q_z  k_z \right) 
 \nonumber\\
&&-\frac{2}{\ell^2}\mathcal{I}_{2}^{n-1,n^\prime-1}\left(q_0^2-q_z^2-q_\perp^2\right) 
+\frac{2}{\ell^2} q_\perp^2 \left( \mathcal{I}_{2}^{n-1,n^\prime-1}-\mathcal{I}_{3}^{n-1,n^\prime-1}\right)  ,  \\
\tilde{S}_2&=&- m^2 \left(\mathcal{I}_{0}^{n,n^\prime}+\mathcal{I}_{0}^{n-1,n^\prime-1}\right) \left(q_0^2-q_z^2-2 q_\perp^2\right) 
- m^2 \left(\mathcal{I}_{0}^{n,n^\prime-1}+\mathcal{I}_{0}^{n-1,n^\prime}\right)  \left(2q_0^2-2q_z^2-q_\perp^2  \right)  ,   \\
\tilde{S}_3&=&2 s_\perp  \left[ (k_z+q_z)E_{n} -k_z E_{n^\prime} \right] \Bigg[
2  \left(\mathcal{I}_{0}^{n,n^\prime}-\mathcal{I}_{0}^{n-1,n^\prime-1}\right)  \left( q_0 E_{n^\prime}-q_z (k_z+q_z) \right)  
- \left(\mathcal{I}_{0}^{n,n^\prime}-\mathcal{I}_{0}^{n-1,n^\prime-1}\right)   \left(q_0^2-q_z^2\right) 
 \nonumber \\
&&+  \left(\mathcal{I}_{0}^{n,n^\prime-1}-\mathcal{I}_{0}^{n-1,n^\prime}\right) 
\left(2q_0^2-2q_z^2 -q_\perp^2\right)  - \frac{2\sqrt{2\xi}}{\ell^2}  \left(\mathcal{I}_{1}^{n,n^\prime-1}-\mathcal{I}_{1}^{n-1,n^\prime-1}\right) 
+\frac{2\sqrt{2\xi}}{\ell^2} \left(\mathcal{I}_{1}^{n^\prime,n-1}-\mathcal{I}_{1}^{n^\prime-1,n-1}\right)  \Bigg] ,
\end{eqnarray}
where, for simplicity of notation, we have suppressed the dependence on $\xi_q$ in the functions $\mathcal{I}_{i}^{n,n^\prime}(\xi_q)$. Additionally, we have used the relations $q_0 k_z- q_zE_{n} =q_0 (k_z+q_z)- q_zE_{n^\prime} =k_z E_{n^\prime}- (k_z+q_z)E_{n} $. The functions $\tilde{S}_i$ can be further simplified by employing the expression for the total neutrino-antineutrino energy, $q_0=E_{n^\prime,k_z+q_z}-E_{n,k_z}$, which is enforced by the $\delta$-function in Eq.~(\ref{res-L-ImPi}). In particular, we find that the following relations hold:
\begin{eqnarray}
E_{n} E_{n^\prime} -k_z (k_z+q_z) &=& \frac{n+n^\prime}{\ell^2}-\frac{q_0^2-q_z^2}{2}+m^2,\\
q_0 E_{n} - q_z  k_z  &=& \frac{n^\prime-n}{\ell^2}-\frac{q_0^2-q_z^2}{2},\\
q_0 E_{n^\prime} - q_z  (k_z+q_z) &=& \frac{n^\prime-n}{\ell^2}+\frac{q_0^2-q_z^2}{2},\\
\left(q_0 k_z- q_zE_{n} \right)^2 &=& \left(\frac{2n}{\ell^2}+m^2\right)q_z^2+\frac{2(n-n^\prime)}{\ell^2}k_zq_z+k_z (k_z+q_z) (q_0^2-q_z^2).
\end{eqnarray}
Using these relations, we derive significantly simpler expressions for $\tilde{S}_i$, i.e., 
\begin{eqnarray}
\tilde{S}_1&=& \left[-2 (q_0^2-q_z^2-q_\perp^2) \frac{n+n^\prime}{\ell^2} 
-m^2\left(q_0^2-q_z^2\right) \right]
\left(\mathcal{I}_{0}^{n,n^\prime} +\mathcal{I}_{0}^{n-1,n^\prime-1}\right) \nonumber\\
&+&\left[2 (q_0^2-q_z^2-q_\perp^2) \frac{n+n^\prime}{\ell^2} 
- \left(q_0^2-q_z^2 -q_\perp^2 \right)^2 
+m^2 \left(2q_0^2-2q_z^2-q_\perp^2\right)
\right] \left(\mathcal{I}_{0}^{n,n^\prime-1} +\mathcal{I}_{0}^{n-1,n^\prime}\right) , 
\label{S1-best-app}  \\
\tilde{S}_2&=&- m^2 \left(\mathcal{I}_{0}^{n,n^\prime}+\mathcal{I}_{0}^{n-1,n^\prime-1}\right) \left(q_0^2-q_z^2-2 q_\perp^2\right) 
- m^2 \left(\mathcal{I}_{0}^{n,n^\prime-1}+\mathcal{I}_{0}^{n-1,n^\prime}\right)  \left(2q_0^2-2q_z^2-q_\perp^2  \right)  ,   
\label{S2-best-app}  \\
\tilde{S}_3&=& 2 s_\perp  \left[ (k_z+q_z)E_{n} -k_z E_{n^\prime} \right] \left[\frac{2 (n^\prime-n)}{\ell^2} \left(\mathcal{I}_{0}^{n,n^\prime}-\mathcal{I}_{0}^{n-1,n^\prime-1}\right) + \left( \mathcal{I}_{0}^{n,n^\prime-1}-\mathcal{I}_{0}^{n-1,n^\prime}\right)  \left(2q_0^2-2q_z^2 - 3 q_\perp^2\right) 
 \right] .
 \label{S3-best-app}
\end{eqnarray}
In the derivation, we have also utilized the following relations among the form factor functions $\mathcal{I}_{i}^{n,n^\prime}\left(\xi_q\right)$ \cite{Wang:2021ebh}:
\begin{eqnarray}
\mathcal{I}_{1}^{n^\prime,n-1}(\xi) -\mathcal{I}_{1}^{n,n^\prime-1}(\xi) &=& \frac{\sqrt{2}}{\sqrt{\xi}} (n-n^\prime) \mathcal{I}_{0}^{n,n^\prime}(\xi) ,\\
\mathcal{I}_{1}^{n^\prime-1,n-1}(\xi) -\mathcal{I}_{1}^{n-1,n^\prime-1}(\xi) &=& \frac{\sqrt{2}}{\sqrt{\xi}} (n-n^\prime) \mathcal{I}_{0}^{n-1,n^\prime-1}(\xi) ,\\
\mathcal{I}_{1}^{n^\prime-1,n-1}(\xi)+\mathcal{I}_{1}^{n,n^\prime-1}(\xi) &=& \sqrt{2\xi} \mathcal{I}_{0}^{n,n^\prime-1}(\xi),\\
\mathcal{I}_{1}^{n^\prime,n-1}(\xi)+\mathcal{I}_{1}^{n-1,n^\prime-1}(\xi)&=& \sqrt{2\xi} \mathcal{I}_{0}^{n-1,n^\prime}(\xi) ,\\
\mathcal{I}_{2}^{n-1,n^\prime-1}(\xi)&=& \frac{n+n^\prime}{2}\left(\mathcal{I}_{0}^{n,n^\prime}(\xi) +\mathcal{I}_{0}^{n-1,n^\prime-1}(\xi)\right)
-\frac{\xi}{2}\left(\mathcal{I}_{0}^{n,n^\prime-1}(\xi) +\mathcal{I}_{0}^{n-1,n^\prime}(\xi)\right),\\
\mathcal{I}_{3}^{n-1,n^\prime-1}(\xi)&=& \frac{n+n^\prime}{2}\left(\mathcal{I}_{0}^{n,n^\prime-1}(\xi) +\mathcal{I}_{0}^{n-1,n^\prime}(\xi)\right)
-\frac{(n-n^\prime)^2}{2\xi} \left(\mathcal{I}_{0}^{n,n^\prime}(\xi) +\mathcal{I}_{0}^{n-1,n^\prime-1}(\xi)\right) .
\end{eqnarray}

In the context of dense quark matter, where the quark chemical potential greatly exceeds other relevant energy scales, an additional approximation can be employed to further simplify the expressions for the functions $\tilde{S}_i$, defined in Eqs.~(\ref{S1-best-app}) through (\ref{S3-best-app}). As discussed in Sec.~\ref{sec:dense-quark-matter}, the presence of a large number of occupied Landau levels allows us to replace the sum over the Landau-level index $n$ with an integral over the transverse quark momentum $k_\perp$, defined via $2n|e_f B| = k_\perp^2$. It is important to emphasize that this $k_\perp$ is distinct from the pseudomomentum $k_\perp$ appearing in the translationally invariant part of the quark propagator in Eqs.~(\ref{prop-quark}) and (\ref{Dn-quark}), which has already been integrated out in Eq.~(\ref{res-L-ImPi}). Applying this approximation yields the following simplified relations:
\begin{eqnarray}
q_0 &\simeq& \frac{k_z q_z +s |e_f B|}{E_n} \simeq q_z \cos\theta +s \delta\epsilon_B,\\
(k_z+q_z) E_{n} - k_z E_{n^\prime}& \simeq & \frac{(k_\perp^2+m^2 ) q_z-s |e_f B| k_z}{E_{n}} \simeq \mu_{f} \left( q_z \sin^2\theta - s\delta\epsilon_B \cos\theta \right).
\end{eqnarray}
Here, by definition, $\delta\epsilon_B \equiv |e_f B|/\mu_{f}$, $s=n^\prime-n$,  $k_z =k \cos\theta$, and $k_\perp =k \sin\theta$. Since the relevant quark momenta $k$  are confined to a narrow region near the Fermi surface, we approximate $k$ by $k_F\approx \mu_f$.

Finally, by taking into account the approximate expressions for the form factor functions in Eq.~(\ref{I00-approx}), we obtain 
\begin{eqnarray}
 \tilde{S}_1&\simeq &
 2 \mu_{f}^2  \left( \left[J_{s-1}\left(a\right)\right]^2+\left[J_{s+1}\left(a\right)\right]^2-2\left[J_{s}\left(a\right)\right]^2\right) (q_0^2-q_z^2-q_\perp^2) \sin^2\theta  +O(m^2 T^2,T^4) ,
\label{S1-approx-app} \\
\tilde{S}_2&=&- 2 m^2 \left(q_0^2-q_z^2-2 q_\perp^2\right) \left[J_{s}\left(a\right)\right]^2
- m^2  \left(2q_0^2-2q_z^2-q_\perp^2  \right) 
\left( \left[J_{s-1}\left(a\right)\right]^2+\left[J_{s+1}\left(a\right)\right]^2\right) \sim O(m^2 T^2) ,   
 \label{S2-approx-app} \\
\tilde{S}_3&=&2 s_\perp \mu_{f} \left( q_z \sin^2\theta - s\delta\epsilon_B \cos\theta \right) \left(2q_0^2-2q_z^2 -q_\perp^2\right) \left( \left[J_{s-1}\left(a\right)\right]^2-\left[J_{s+1}\left(a\right)\right]^2\right) \sim O(\mu_{f} T^3),
 \label{S3-approx-app}
\end{eqnarray}
where $a\equiv k_\perp q_\perp \ell^2 \approx \mu_{f} q_\perp \ell^2 \sin\theta$. It is important to note that the functions $\tilde{S}_2$ and $\tilde{S}_3$ give subleading contributions, suppressed by powers of $\mu_f$ relative to $ \tilde{S}_1$.

%\bibliography{synchrotron}


\end{document}